\documentclass[11pt]{article}
\usepackage{amsmath,amsbsy,amsfonts,amssymb,graphics,epsfig,float,mathrsfs,amsthm,braket}
\usepackage{ifsym}
\usepackage{psfrag}
\usepackage{color}
\usepackage[normalem]{ulem}

\newcommand \beq{\begin{equation}}
\newcommand \eeq{\end{equation}}
\newcommand \beqn{\begin{equation*}}
\newcommand \eeqn{\end{equation*}}

\newcommand{\wo}{\Delta}

\newcommand{\upd}{\mathrm{d}}

\newcommand{\atanh}{\mathrm{arctanh}}

\newcommand{\sqq}{\sigma_{\theta\theta}}
\newcommand{\srr}{\sigma_{rr}}

\newcommand{\tf}{\tilde{f}}
\newcommand{\tchi}{\tilde{\chi}}
\newcommand{\trho}{\tilde{\rho}}
\newcommand{\tphi}{\tilde{\Phi}}
\newcommand{\hf}{\widehat{f}}
\newcommand{\hchi}{\widehat{\chi}}
\newcommand{\hrho}{\widehat{\rho}}

\newcommand{\cpogo}{c_{\mathrm{pogo}}}

\newcommand{\s}[1]{{\textsf{\textbf{#1}}}}

\begin{document}
\title{\s{The shallow shell approach to Pogorelov's problem and the breakdown of `mirror buckling'}}
\author{ \textsf{Michael Gomez$^\dagger$, Derek E. Moulton$^\dagger$ and Dominic Vella$^\dagger$}\\ 
{\it$^\dagger$Mathematical Institute, University of Oxford, UK}}

\date{\today}
\maketitle
\hrule\vskip 6pt
\begin{abstract}
We present a detailed asymptotic analysis of the point indentation of an unpressurized, spherical elastic shell. Previous analyses of this classic problem have assumed that for sufficiently large indentation depths, such a shell deforms by `mirror buckling' --- a portion of the shell inverts to become a spherical cap with equal but opposite curvature to the undeformed shell. The energy of deformation is then localized in a ridge in which the deformed and undeformed portions of the shell join together, commonly referred to as  Pogorelov's ridge. Rather than using an energy formulation, we revisit this problem from the point of view of the shallow shell equations and perform an asymptotic analysis that exploits the largeness of the indentation depth. This reveals first that the stress profile associated with mirror buckling is singular as the indenter is approached. This consequence of point indentation means that mirror buckling must be modified to incorporate the shell's bending stiffness close to the indenter and gives rise to an intricate asymptotic structure with seven different spatial regions. This is in contrast with the three regions (mirror-buckled, ridge and undeformed) that are usually assumed and yields new insight into the large compressive hoop stress that ultimately causes the shell to buckle azimuthally.
\end{abstract}
\vskip 6pt
\hrule

\maketitle

%%%%%%%%%%%%%%% End of first page %%%%%%%%%%%%%%%%%%%%%

\section{Introduction}
When a thin, spherical shell is `poked', everyday experience suggests that the displacement of the shell is restricted to a `dimple' around the indentation point. As the magnitude of the imposed indentation becomes large compared to the thickness of the shell, detailed analysis shows that this dimple becomes arbitrarily close to that of an inverted spherical cap~\cite{pogorelov}.  This `mirror-buckled' solution is classic in the literature and is often used as  the  simplest geometrical ingredient with which to understand the deformation of shells \cite{vella12,xu14}, since it is also observed in other loading settings such as a shell pushed by a plane~\cite{audoly10} or compression of a shell by an external pressurization~\cite{knoche2014}.

\begin{figure}[!htb]
\centering
\includegraphics[width=12cm]{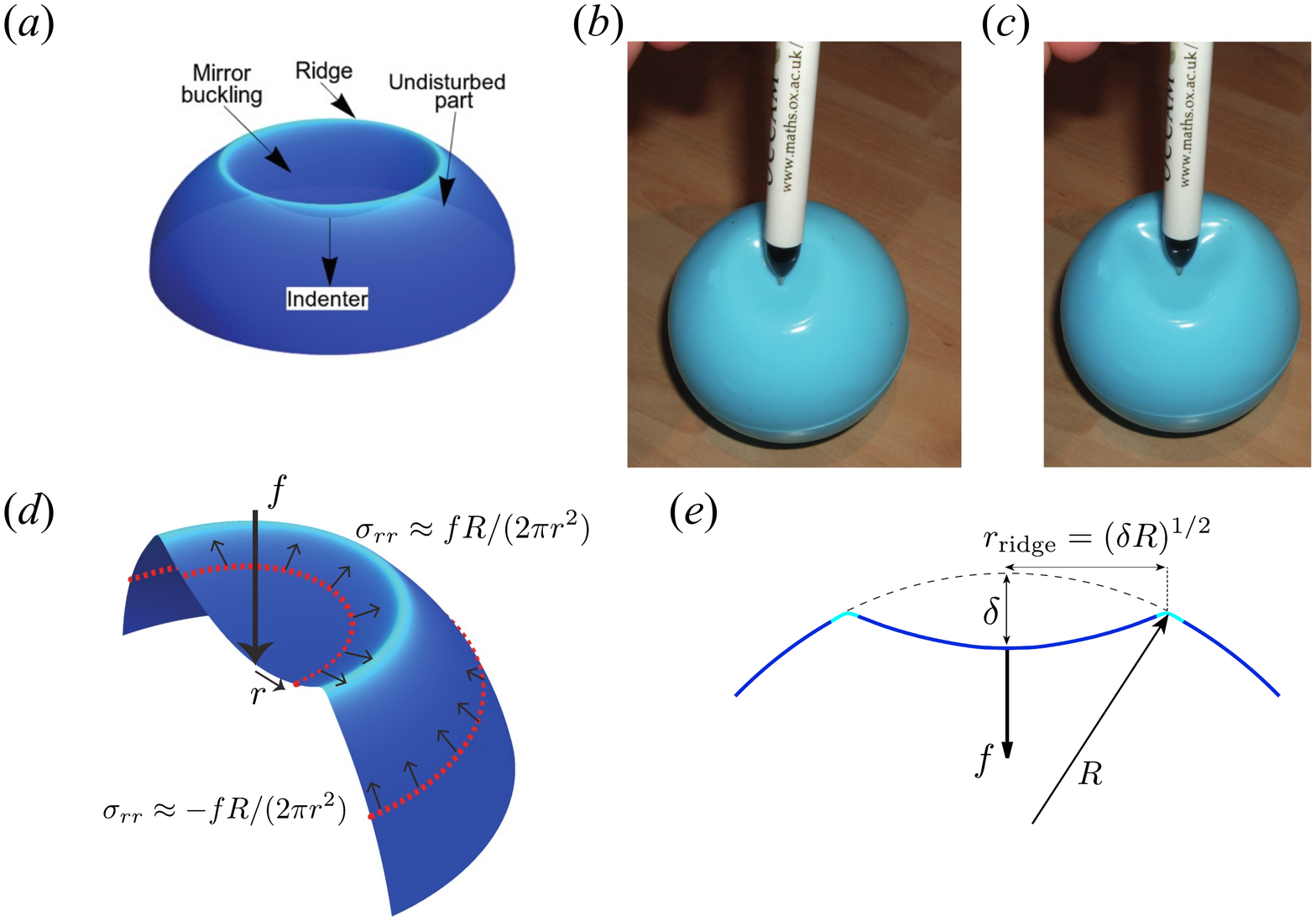}
\caption{(a) The classic three-region structure describing the deformation of a thin, spherical shell for large indentations. The deformation is localized to an inverted spherical cap that is smoothly connected to the undisturbed region by a narrow ridge (lighter shading); this neglects the role of bending effects near the point of indentation. (b) At moderate indentations a shell remains axisymmetric; (c) at larger indentations axisymmetry is lost through secondary azimuthal buckling. (d) The direction of meridional stress acting on parts of the shell inside and outside the dimple, in relation to the indentation force. (e) The (dimensional) geometrical parameters of the dimple under mirror buckling.}
\label{mirrorbuckling}
\end{figure}

The ubiquity of mirror buckling arises from the fact that it is an isometric (distance preserving) deformation of the original sphere in all but a small ridge region where the inverted cap connects to the remainder of the sphere (see figure \ref{mirrorbuckling}a). The energy contained within this ridge is determined by a balance between bending and stretching effects but, crucially, vanishes as the thickness, $h$, of the shell vanishes \cite{landau} (in fact, the energy $\sim h^{5/2}$ as $h \to 0$~\cite{audoly10}). For a small but finite thickness, the dominant role of inextensional bending summarized in Love's `principle of applicable surfaces'~\cite{love1927} dictates that this remains the energetically favourable configuration. 

Previous analytical studies of the point-loading problem for large indentations have focused on understanding the inherently nonlinear behaviour of the solution, including the properties of the ridge layer \cite{ashwell1959,ranjan1977,pogorelov,libai,landau,audoly10}. This is of fundamental interest for engineering applications because it allows a calculation of the force-displacement relation, extending the small-indentation result of Reissner \cite{reissner47a}. These works are all predicated on a three-region deformation structure, namely the ridge layer and the two (oppositely curved) spherical parts (figure~\ref{mirrorbuckling}a). These authors have therefore assumed that the mirror-buckled solution can be taken to be a global description of the deformation of the shell. In fact, for a deformation caused by indentation, the mirror-buckled solution is singular as the indenter is approached --- a singularity that is resolved by the re-emergence of bending effects close to the indenter. Whilst this aspect has been briefly noted by some authors~\cite{libai,steele1989,vaziri08}, a detailed investigation of the local behaviour, together with its consequences for mirror buckling, is still lacking. 

One further assumption that is made in standard analyses (as well as this one) is that the deformation remains axisymmetric. However, it is known that the deformation eventually loses its axisymmetry \cite{fitch1968} undergoing a secondary buckling instability to form several connected lobes~\cite{penning1965} (see figures \ref{mirrorbuckling}b,c); further lobes appear as indentation progresses \cite{vaziri08}. This buckling is considered to be the initial stage of failure in many engineering applications~\cite{steele1989}, and previous work has focused on quantifying its onset using numerical simulations and describing the post-buckled shapes~\cite{fitch1968,vaziri08,vaziri09}. However, an explanation of the origin of the instability remains elusive. Some understanding has recently been provided by~\cite{knoche2014}, where the dimpled configuration is shown to exhibit a region of compressive circumferential stress along the Pogorelov ridge, essentially due to an inward radial displacement associated with deformation. This compressive stress is strong in the sense that it is able to overcome the threshold for azimuthal buckling as a way for the shell to release the compression. Here, we shed new light on the origin of this strong compression for the case of point indentation. It has also been suggested recently that, even though the mirror buckling isometry is only observed under certain conditions, other `wrinkly isometries'~\cite{vella15EPL} may exist. It is therefore important to better understand the simplest isometry of all, mirror buckling. 

In this paper we explore the indentation of an axisymmetric spherical shell, focusing on the asymptotic structure that emerges when the indentation depth is large compared to the thickness. While a true shell loses axisymmetry for moderately large indentations, our asymptotic analysis of the axisymmetric problem provides new insights into this loss of axisymmetry. We begin in \S\ref{sectiongoverning} by describing the equations governing axisymmetric deformations of a shallow elastic shell, before presenting their numerical solution in \S\ref{sectionstructure}. The numerical results hint at a rich asymptotic structure composed of seven regions in total, with the shell transitioning between states dominated by pure bending or stretching effects. The analysis of this structure forms the focus for the remainder of the paper. The ridge layer is explored in \S\ref{interiorlayer},  where we show that the compressive stress induced by indentation is significantly larger than is actually required to balance the applied force. In \S\ref{solnstructure} we show that mirror buckling breaks down far from both the ridge and the indenter. In this way, we build up a picture of the complete asymptotic structure, which we summarize in figure~\ref{asympstructure}. Finally we derive the corresponding force-displacement relationship, \S\ref{forcelaw}, before discussing our findings in \S\ref{sectionconclusion}.

\section{Theoretical formulation}
\label{sectiongoverning}

\subsection{Governing equations}

We consider a linearly elastic, spherical shell of radius $R$, thickness $h$, Young's modulus $E$ and Poisson ratio $\nu$. The shell is subject to an indentation $\delta$ at its pole by a point force of magnitude $f$. The radial coordinate (measured horizontally from the pole) is $r$ and the normal displacement of the shell from its undisturbed position is the unknown function $w(r)$ (assuming axisymmetric deformations).

To determine $w(r)$ for a given force $f$, we will use the so-called `shallow shell' formulation~\cite{ventsel01}. This assumes that the horizontal scale of the deformation is significantly smaller than the radius of the shell $R$; for larger deformations the  shell is said to be `deep'~\cite{libai}. In the shallow shell formulation, the middle surface of the shell is  effectively represented by a paraboloid of revolution. In this way, the equilibrium equations describing axisymmetric, torsionless deformations of an isotropic, homogeneous shell of small thickness $h \ll R$ can be greatly simplified \cite{ventsel01}. In particular, vertical force-balance reads
\begin{equation}
B\nabla^4w+\frac{1}{R}\frac{1}{r}\frac{\upd }{\upd r}\left(r\psi\right)-\frac{1}{r}\frac{\upd }{\upd r}\left(\psi\frac{\upd w}{\upd r}\right)=-\frac{f}{2\pi}\frac{\delta(r)}{r},
\label{fvk1:dim}
\end{equation} where $\delta(r)$ is the Dirac $\delta$-function, while the equation expressing the compatibility of strains reads
\begin{equation}
\frac{1}{Eh}r\frac{\upd }{\upd r}\left[\frac{1}{r}\frac{\upd }{\upd r}\left(r\psi\right)\right]=\frac{r}{R}\frac{\upd w}{\upd r}-\frac{1}{2}\left(\frac{\upd w}{\upd r}\right)^2.
\label{fvk2:dim}
\end{equation} Here $B=Eh^3/[12(1-\nu^2)]$ is the bending stiffness of the shell and $\psi$ is the derivative of the Airy stress function, so that the in-plane stresses are $\sqq=\psi'$ and $\srr=\psi/r$; the corresponding radial displacement is $u=(r \psi'-\nu \psi)/(Eh)$. It is important to note that these equations are valid only under the assumption of small slopes $|\upd w/\upd r|\ll1$ (see the discussion in \cite{audoly10}, for example).

Equations \eqref{fvk1:dim}--\eqref{fvk2:dim} are well-known and can be found in many textbooks (see \cite{calladine83,bazant91,ventsel01}, for example\footnote{Letting $R \to \infty$, one recovers the perhaps more familiar F\"{o}ppl-von-K\'{a}rm\'{a}n equations, which describe axisymmetric deformations of a naturally flat plate.}). 
For ease of reference to the various terms in \eqref{fvk1:dim}--\eqref{fvk2:dim}, we briefly discuss their physical origins. The $B \nabla^4w$ term represents the net transverse shear force due to bending of the mid-surface, while the nonlinear terms $r^{-1}\upd\left(\psi\,\upd w/\upd r\right)/\upd r$ and $\left(\upd w/\upd r\right)^2$
%$r^{-1}\upd\left(\psi~\upd w/\upd r\right)/\upd r$ and $\left(\upd w/\upd r\right)^2$ 
account for the lowest geometric nonlinearities coupling out-of-plane rotations to in-plane membrane strains. We therefore refer to these as the rotation terms. The second equation expresses compatibility of in-plane strains, with the right-hand side being the change in Gauss curvature due to deformation. The additional terms in \eqref{fvk1:dim}--\eqref{fvk2:dim} for finite $R$ then correspond to modifications due to the finite radius of curvature of the undeformed shell; we refer to these terms as the curvature terms. %In the absence of any displacement about the spherical shape, \eqref{fvk1:dim} is entirely analogous to the Young-Laplace equation for a capillary surface: the non-zero curvature allows the in-plane stresses to support transverse forces.

\subsection{Non-dimensionalization}
We non-dimensionalize the governing equations \eqref{fvk1:dim}--\eqref{fvk2:dim} by noting that a balance between bending and stretching leads to the emergence of a radial scale $(h R)^{1/2}$ \cite{reissner47a,vella12}; we therefore let
\begin{equation}
\rho=r/(hR)^{1/2},\quad W=w/h,\quad \Psi=\frac{\psi}{Eh^2}\left(\frac{R}{h}\right)^{1/2},\quad F = \frac{f R}{E h^3}.
\label{p0:ndim}
\end{equation} We obtain
\begin{equation}
\frac{1}{12(1-\nu^2)}\nabla^4W+\frac{1}{\rho}\frac{\upd }{\upd \rho}\left(\rho\Psi\right)-\frac{1}{\rho}\frac{\upd }{\upd \rho}\left(\Psi\frac{\upd W}{\upd \rho}\right)=-\frac{F}{2 \pi}\frac{\delta(\rho)}{\rho},
\label{fvk1:nd}
\end{equation} and
\begin{equation}
\rho\frac{\upd }{\upd \rho}\left[\frac{1}{\rho}\frac{\upd }{\upd \rho}\left(\rho\Psi\right)\right]=\rho\frac{\upd W}{\upd\rho}-\frac{1}{2}\left(\frac{\upd W}{\upd \rho}\right)^2,
\label{fvk2:nd}
\end{equation} which are to be solved with the boundary conditions
\beq
W(0)=-\delta/h=-\wo,\quad W'(0)=0,\quad\lim_{\rho\to0}\left[\rho\Psi'(\rho)-\nu\Psi(\rho)\right]=0,
\label{nd:bcs0}
\eeq and 
\beq
W(\infty)=W'(\infty)=\Psi(\infty)=0.
\label{nd:bcsinf}
\eeq Here $\wo=\delta/h$ is introduced as the dimensionless indentation depth. It is mathematically simpler to impose a given indentation depth at the origin and then calculate the force $F$ required to impose this indentation \emph{a posteriori}. The second condition in \eqref{nd:bcs0} specifies that the vertical displacement has zero slope at the indentation point, and the third condition corresponds to zero radial displacement at $r=0$; together these eliminate the possibility of a cusp or tear in the shell at the indenter. The conditions in \eqref{nd:bcsinf} state that far from the indentation point, the shell must tend back to its undisturbed state.

As a result of our non-dimensionalization,  $\wo$ emerges as the only control parameter in the problem; our analysis will focus on understanding the behaviour of the system in the limit of $\wo\gg1$. Unless stated otherwise, this is the limit we are considering, e.g.~we write $f = O(g)$ for $f = O(g)$ as $\Delta\to\infty$. The Poisson ratio $\nu$ is the other dimensionless parameter in the problem but is constant during an indentation experiment. Furthermore, it is often convenient to write $$k=[12(1-\nu^2)]^{1/4}.$$ While it is possible to scale out the $k^{-4}$ factor in \eqref{fvk1:nd} by using the alternative radial scale $(h R)^{1/2} k^{-1}$ in the non-dimensionalization~\cite{libai}, our non-dimensionalization has the advantage that the vertical scale is simply $h$: $\wo$ is simply the indentation depth relative to the shell thickness.

\section{Structure of the deformation}
\label{sectionstructure}

\subsection{Undeformed solution\label{sec:flat}}
In the loading regime under consideration, the key observation is that the displacement of the shell is localized to a dimple around the indentation point. Outside of this neighbourhood the shell approaches its stress-free, undeformed state: $W,\Psi\to0$. For sufficiently large $\rho$, therefore, $W,\Psi\ll1$ and we may linearize the governing equations. The linearized compatibility equation \eqref{fvk2:nd} can be immediately integrated once to find that
\beq
\frac{1}{\rho}\frac{\upd }{\upd \rho}\left(\rho\Psi\right)=W \label{nd:flat1},
\eeq
where the constant of integration is zero by the far-field conditions \eqref{nd:bcsinf}. Substituting into \eqref{fvk1:nd} and linearizing gives 
\beq
k^{-4}\nabla^4 W + W = 0 \label{nd:flat2},
\eeq
where the point force does not enter away from $\rho = 0$. The general solution decaying at infinity can be written as a linear combination of the Kelvin functions $\mathrm{ker}(k\rho)$ and $\mathrm{kei}(k\rho)$~\cite{olver2010}. It follows that $W$ is exponentially small in $\rho$, from which \eqref{nd:flat1} gives that $\Psi \propto 1/\rho$ to leading order.

The $1/\rho$ behaviour appearing in the solution for $\Psi$ can also be derived heuristically by considering force-balance within the shell. At any radius beyond the Pogorelov ridge, the resultant of the dimensional indentation force $f$ must be balanced by the vertical component of the meridional stress $\srr$, integrated around the circle; see figure~\ref{mirrorbuckling}d. This gives that
\beqn
 2\pi r\srr \cdot \frac{r}{R} \sim -f \quad \mathrm{as}\quad r\to\infty,
\eeqn
with the $r/R$ factor arising from the (small) angle made by the mid-surface of the shell to the vertical. In dimensionless terms, it follows that
\beq W = o(\rho^n)\quad \forall n, \quad \Psi\sim -\frac{F}{2\pi\rho}\quad \mathrm{as}\quad \rho\to\infty. \label{flatsoln}
\eeq We refer to the solution in \eqref{flatsoln} as the `undeformed' solution since the displacement of the shell is exponentially small in this region. 

The nonlinear terms that were neglected in \eqref{nd:flat1}--\eqref{nd:flat2} involve $W$ and its derivatives and must, because of \eqref{flatsoln}, also be exponentially small in $\rho$. As we shall see below, the position of the ridge layer is at $\rho=O(\Delta^{1/2})$ as $\Delta \to \infty$. The undeformed solution \eqref{flatsoln} is then valid for $\rho \gg \Delta^{1/2}$. Since $W$ is exponentially small in $\rho$ ($\gg \Delta^{1/2}$), it must also be exponentially small in $\wo$, together with any higher-order terms in the expansion, and we conclude that \eqref{flatsoln} is  correct to all powers of $\Delta$. (Alternatively, expanding $W \sim \Delta^{-n}W_n(\xi)$ where $\xi \equiv \Delta^{-1/2}\rho \gg 1$ shows that $W_n = 0$ for each $n > 0$.)

Before proceeding, note that we can integrate the force-balance \eqref{fvk1:nd} from infinity down to $\rho$ to obtain a form that is more convenient for analysis; the resulting integration constant is obtained from \eqref{flatsoln} and gives
\beq
k^{-4} \rho\frac{\upd }{\upd \rho}\left[\frac{1}{\rho}\frac{\upd }{\upd \rho}\left(\rho\frac{\upd W}{\upd \rho}\right)\right] + \rho\Psi-\Psi\frac{\upd W}{\upd \rho} = -\frac{F}{2\pi}. \label{fvk1int:nd}
\eeq Note that \eqref{fvk1int:nd} is valid for all $\rho$.

%\noteDV{Do we really need this?}Note that the first term in \eqref{fvk1int:nd} represents bending forces, whilst the dimensionless rotation terms have become $\Psi W'$ together with $W'^2$ in \eqref{fvk2:nd}. The dimensionless curvature terms are $\rho\Psi$ and $\rho W'$, so that the flat solution can be re-interpreted as a balance between the rest curvature and the force resultant in \eqref{fvk1int:nd}. 

\subsection{Mirror-buckled solution\label{sec:mirror}}

The geometrical idea behind `mirror-buckling' is to search for an isometric  deformation free of stretching stresses to leading order in $1/\Delta$. In particular there can be no change in the Gauss curvature of the mid-surface, i.e.~the right-hand side of the compatibility equation \eqref{fvk2:nd} is zero: 
\beqn
\rho\frac{\upd W}{\upd\rho}-\frac{1}{2}\left(\frac{\upd W}{\upd \rho}\right)^2 = 0.
\eeqn
One solution is the undeformed solution discussed above, $W' = 0$. The other solution is $W' = 2 \rho$; substituting into the integrated vertical force balance \eqref{fvk1int:nd} yields
\beq
W \sim A_1+\rho^2, \quad \Psi \sim \frac{F}{2 \pi\rho}. \label{mirrorsoln}
\eeq
Here $A_1$ is an (as yet undetermined) constant. We have used $\sim$ to emphasize that \eqref{mirrorsoln} is only an asymptotic solution of the original system. In particular, this solution cannot satisfy the boundary conditions at the indenter at the origin (since $\Psi$ would be singular) or in the far-field (since $W$ would diverge) --- it can only hold in some intermediate region.

Note that the vertical displacement $W$ in \eqref{mirrorsoln} represents, to within the shallowness approximation, a reflected spherical cap, i.e.~the mirror-buckled solution. Furthermore, note that the sign of $\Psi$ has changed  compared to \eqref{flatsoln}: the curvature of the mirror-buckled region is in the opposite sense to the undeformed shell so, to balance the indentation force, the meridional stress within the shell must be tensile ($\srr>0$), rather than compressive ($\srr<0$); see figure~\ref{mirrorbuckling}d.

At this point, it is tempting to  choose $A_1=-\Delta$ to satisfy $W(0)=-\Delta$. However, given the earlier observation that mirror-buckling can only hold in an intermediate region, our neglect of bending effects must be invalid sufficiently close to the indenter. Nevertheless, we expect  $A_1=-\Delta$ to \emph{leading-order}: any correction to the displacement arising from the breakdown of  mirror-buckling must be small compared to $\Delta$, an assumption that will be explicitly verified later.

With $A_1 \sim -\Delta$, we see that the mirror-buckled solution \eqref{mirrorsoln} patches onto the undeformed solution, $W = o(\Delta^n)~\forall n$, when $\rho^2$ becomes comparable to $\Delta$, so that the radius of the ridge layer is at $O(\Delta^{1/2})$. This can be confirmed by an order-of-magnitude estimate based on the geometry, which we know to be close to a reflected spherical cap: from figure~\ref{mirrorbuckling}e, simple trigonometry gives  $r_{\mathrm{ridge}}^2 \approx \delta R$. This gives that $\rho \approx \Delta^{1/2}$, which is valid provided $\delta \ll R$ (a shallow deformation). 

\subsection{Numerical solution of the full problem}

Since the conventional three-region view (mirror-buckling, ridge and undeformed) requires a singular stress field as $\rho \to 0$, this picture cannot represent the full story for point indentation close to the indenter. It is therefore useful to consider the numerical solution of the full problem \eqref{fvk1:nd}--\eqref{nd:bcsinf}. This is easily accomplished  using the \textsc{matlab} routine \texttt{bvp4c} on a finite domain that excludes the origin to avoid a coordinate singularity there\footnote{We used $[5\times10^{-6},10^4]$ as our computational domain with the boundary condition at $\rho_0=5\times10^{-6}$ modified by the appropriate  asymptotic results for $\rho\ll1$ discussed in \S\ref{solnstructure}.}. The deformation and stress function caused by an imposed indentation depth $\wo$ can be computed in this manner and the force $F$ required to impose this indentation is determined from \eqref{fvk1int:nd}.

\begin{figure}
\centering
\includegraphics[width=12cm]{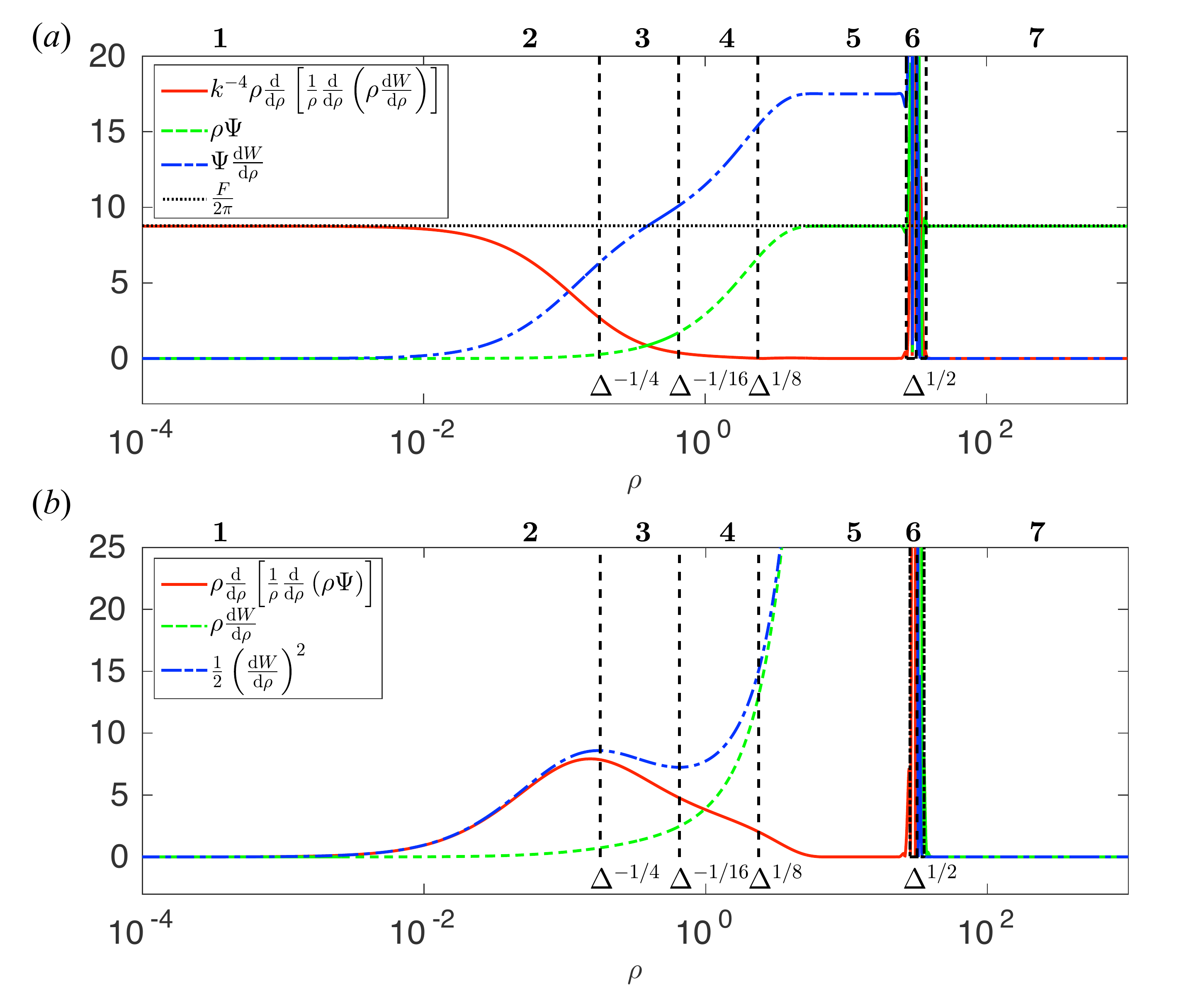}
\caption{The individual terms in the force-balance equation \eqref{fvk1int:nd} (top) and the compatibility equation \eqref{fvk2:nd} (bottom) determined from the numerical solution of the full dimensionless problem \eqref{fvk1:nd}--\eqref{nd:bcsinf}. In each case the absolute value is plotted. The seven regions corresponding to different sets of balances in the equations are labelled. For later reference, the scaling behaviour of the boundaries between these regions are also shown (vertical dashed lines). (Note that the use of logarithmic axes over-emphasizes the size of the inner regions $1$--$4$.) Here the indentation is $\Delta = 10^3$ and $\nu = 0.3$.}%The left-hand boundary in the solver is taken to be $\rho_0 = 5 \times 10^{-6}$.}
\label{sizeofterms1}
\end{figure}

\noindent
 
Figure~\ref{sizeofterms1} displays the individual terms of the (integrated) force-balance equation \eqref{fvk1int:nd} and the compatibility equation \eqref{fvk2:nd} as functions of $\rho$, plotted on a semi-logarithmic scale for $\wo=10^3$. Plotted in this way, the different balances between terms in the equations become explicit. 

\begin{figure}
\centering
\includegraphics[width=12cm]{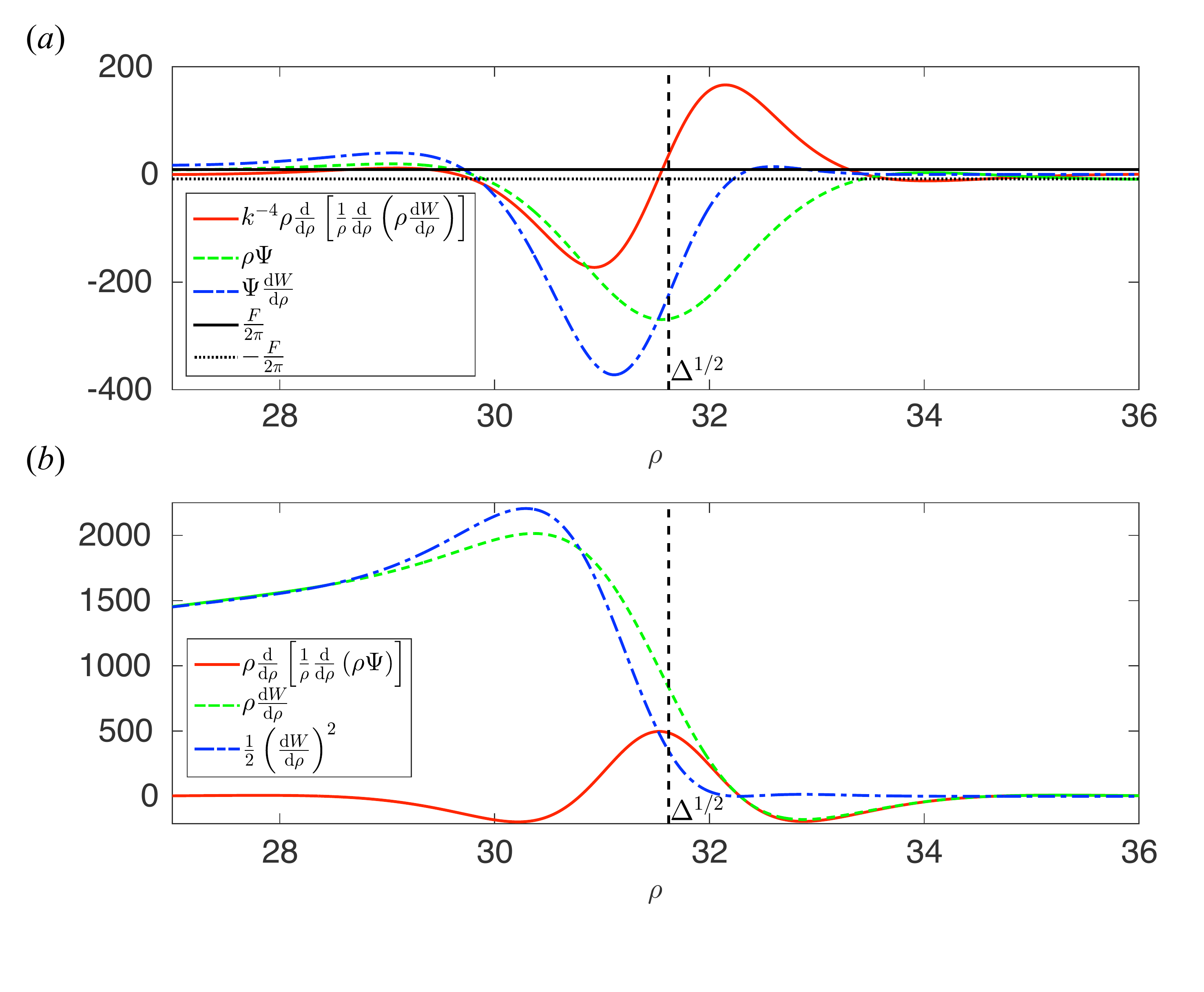}
\caption{A close-up of the ridge layer (region 6 in figure \ref{sizeofterms1}), showing individual terms in the force-balance equation \eqref{fvk1int:nd} (top) and the compatibility equation \eqref{fvk2:nd} (bottom) for $\Delta = 10^3$ and $\nu = 0.3$. The region displayed is that contained in the dash-dotted box of figure~\ref{sizeofterms1}; at its edges are the transitions to the undeformed and mirror-buckled solutions.}
\label{sizeofterms2}
\end{figure}
\noindent

We see from figure~\ref{sizeofterms1} that there are a number of different balances between terms as $\rho$ varies; we shall argue that there are seven different regions in total. The outer, undeformed solution applies far from the indenter, in region $7$: the force-balance here reduces to a balance between the force resultant $F/(2 \pi)$ and the curvature term $\rho \Psi$, with other terms negligible in comparison (as expected from \S\ref{sectionstructure}\ref{sec:flat}). Mirror buckling occurs in region $5$, where stretching effects dominate bending: both of the rotation terms $\Psi W'$ and $W'^2$ are important in this region.

The boundary between regions $5$ and $7$, region $6$, is the classic Pogorelov ridge layer in which  all terms are in balance due to the importance of both bending and stretching effects. A zoom of this region on linear axes is displayed in figure~\ref{sizeofterms2} and shows that, in transitioning between the behaviours expected in the mirror-buckled and undeformed regions, the stress function needs only to change from $\Psi \sim F/(2 \pi\rho)$ to $\Psi \sim -F/(2 \pi\rho)$; however, in making this change of sign, it in fact overshoots by several orders of magnitude. This unexpected overshoot leads to a large, compressive circumferential or `hoop' stress within the shell, since $\sqq = \psi' \propto \Psi'$. This compressive hoop stress in the ridge is thought to be the cause of the secondary buckling instability of a shell \cite{knoche2014} and so understanding its origin is a main goal of this paper.

The behaviour exhibited in regions $5$--$7$ is largely the classic picture of mirror buckling offered by Pogorelov \cite{pogorelov} and others. However, moving towards the indenter from region $5$, figure \ref{sizeofterms1} reveals a rich variety of dominant balances: as already noted, the bending term must play a role close to the indenter (to avoid a stress singularity). This expectation is confirmed (region $1$ in figure \ref{sizeofterms1}) but, rather surprisingly, three other regions (regions $2$--$4$ in figure \ref{sizeofterms1}) are needed to connect this to the classic mirror buckling solution. The existence of these regions has not, to our knowledge, been reported previously. We also note that the need for these additional regions will not occur in other scenarios. For example, in the indentation of a shell by a  plane,  contact occurs only within the Pogorelov ridge (our region 6) and the mirror buckling solution is valid everywhere within the inner region~\cite{audoly10}. The complication in indentation arises because of the singular stress field that would be needed to balance a constant vertical force at $\rho=0$ (where the shell is horizontal); a transverse force due to bending is needed to achieve equilibrium.

While the existence of these additional regions is surprising, the different balances that hold within them is perhaps more surprising still: bending is only important in region $2$ (but not $3$ and $4$) while in region $3$ neither the bending stiffness nor the natural curvature of the shell play a role. In region $3$ the shell essentially behaves like a naturally flat membrane!

\section{Ridge layer analysis (region $6$)}
\label{interiorlayer}
To explain the rich behaviour hinted at in figures~\ref{sizeofterms1}--\ref{sizeofterms2}, we first analyse the ridge layer (region $6$). In this region, the solution transitions between the mirror-buckled and undeformed solutions found previously, being also characterized by a balance between bending and stretching with moderate rotations \cite{audoly10}.  Recall that in the ridge $\rho=O(\Delta^{1/2})$, and denote the ridge's width by $\lambda$ where $\lambda \ll \Delta^{1/2}$. Based on the solutions \eqref{flatsoln} and \eqref{mirrorsoln} we would predict that $\Psi$ remains of order $F \Delta^{-1/2}$ throughout the ridge; the $\Delta^{1/2}$ ratio between  $\Psi$ and $F$ here is needed to balance the vertical component of the meridional stress (integrated around the ridge of radius $\Delta^{1/2}$) with the indentation force. If $[W]$ is the typical size of the displacement, then the terms on the right-hand side of \eqref{fvk2:nd} are of magnitude $\Delta^{1/2}[W]\lambda^{-1}$ and $[W]^2\lambda^{-2}$. To account for moderate rotations, we balance these with the remaining term to give that $[W] =O(\lambda\Delta^{1/2})$ and $F =O(\lambda^2\Delta)$; meanwhile, balancing the dominant part of the bending term in \eqref{fvk1int:nd} with the remaining terms of order $F$ gives the orders of magnitude in the ridge as
\beq
\lambda = O(1),\quad F = O(\Delta), \quad W = O(\Delta^{1/2}), \quad \Psi = O(\Delta^{1/2}).
\label{eqn:ridgescale}
\eeq

This formal balance argument overestimates the magnitude of the force, which for large indentations is known to be $O(\Delta^{1/2})$, i.e.~the force is proportional to the \emph{square root} of the indentation depth~\cite{pogorelov}. Contrast this with the picture for small indentations for which the displacement of the shell is confined to an $O(1)$ neighbourhood of the indenter and the force scales linearly with the indentation depth \cite{reissner47a}. The vanishing of a leading-order force for large indentations is a result of the shell accessing the isometry of mirror-buckling, while the stress is focussed into a narrow ridge region. This allows the shell to more efficiently minimize its energy, as seen in related problems~\cite{audoly10}.

Nevertheless, we emphasize that the above overestimate of the force leads to the correct (in the sense of agreeing with numerics) magnitude of the stress function in the ridge, $\Psi = O(\Delta^{1/2})$; taking $F = O(\Delta^{1/2})$ would lead to the (erroneous) prediction $\Psi = O(1)$. The question is then not ``why is $\Psi$ so large?" but rather ``why is the indentation force so small?". To answer this, we turn to a more detailed asymptotic analysis of the ridge.
%%his is an early sign that simple balance arguments can lead to erroneous conclusions, whereas formal asymptotic arguments might account for the interesting behaviour observed in figure~\ref{sizeofterms2}.

We introduce a new coordinate $x = \rho - \wo^{1/2}$ centred on the interior layer and varying on an $O(1)$ length scale. Written in terms of this local variable, the behaviour in the neighbouring mirror-buckled region may be written
\beqn
\frac{\upd W}{\upd x} \sim 2 (\Delta^{1/2} + x), \quad \Psi \sim \frac{F}{2 \pi (\Delta^{1/2}+x)},
\eeqn
while that in the undeformed region is
\beqn
\frac{\upd W}{\upd x} = o(\Delta^n)\quad \forall n, \quad \Psi \sim -\frac{F}{2 \pi (\Delta^{1/2}+x)}.
\eeqn
For convenience we introduce the functions $f(x)$ and $\chi(x)$, defined in the ridge layer by
%\beqn
%W=g(x)+\begin{cases}
%(A_1 + \Delta)+2 x \Delta^{1/2} + x^2,\quad x<0,\\
%0,\quad x>0.
%\end{cases}
%\eeqn
\beq
\frac{\upd W}{\upd x}=f(x)+\begin{cases}
2(\Delta^{1/2} + x), \quad x<0,\\
0,\quad x>0,
\end{cases} \label{defnf}
\eeq
and
\beq
\Psi=\chi(x)+\begin{cases}
\frac{F}{2\pi(\Delta^{1/2}+x)},\quad x<0,\\
-\frac{F}{2\pi(\Delta^{1/2}+x)},\quad x>0. 
\end{cases} \label{defnchi}
\eeq
With $f$ and $\chi$ defined in this way, matching onto the undeformed and mirror-buckled solutions requires only that $f,\chi\to0$ as $|x| \rightarrow \infty$. It can be shown that the undeformed and mirror-buckled  solutions, \eqref{flatsoln} and \eqref{mirrorsoln}, are correct to all orders in $\Delta$. Hence the matching conditions for $f$ and $\chi$ require decay at all orders in $\Delta$. The definitions of $f(x)$ and $\chi(x)$ are such that these functions are discontinuous at $x=0$ (since $W'$, $\Psi$ and their first derivatives remain continuous). In particular, denoting  the jump in $\cdot$ across $x = 0$ by $[\cdot]^{+}_{-}$  we have
\beq
[f]_{-}^{+}=2\wo^{1/2}, \quad [f']_{-}^{+}=2, \quad [\chi]_{-}^{+}=\frac{F}{\pi \Delta^{1/2}}, \quad [\chi']_{-}^{+}=-\frac{F}{\pi \Delta}.
\label{eqn:chidiscty}
\eeq

%Note that the governing equations \eqref{fvk2:nd} and \eqref{fvk1int:nd} do not involve $W$ directly, but rather $W'$. It is convenient to work with the function $f$ instead defined by
%\beq
%\frac{\upd W}{\upd \rho}=f(x)+\begin{cases}
%2(\Delta^{1/2} + x), \quad x<0,\\
%0,\quad x>0,
%\end{cases} \label{defnf}
%\eeq
%so that $g'(x) = f(x)$ for $x < 0$ and $x > 0$ separately. The jump conditions on $f$ and $f'$ are then
%\beq
%[f]_{-}^{+}=2\wo^{1/2}, \quad [f']_{-}^{+}=2.
%\label{eqn:fdiscty}
%\eeq
%For later reference we note that the first jump condition on $g$ in \eqref{eqn:chidiscty} becomes an integral condition on $f$. Since we have that
%\beqn
%g(x)=\begin{cases}
%\int_{-\infty}^{x} f(t) \upd t, \quad x<0,\\
%-\int_{x}^{\infty} f(t) \upd t, \quad x>0,
%\end{cases}
%\eeqn 
%this condition may be written as
%\beq
%A_1 + \Delta = -\left(\int_{-\infty}^{0-}+\int_{0+}^{\infty} \right) f(t) \upd t. \label{integralcondition}
%\eeq
 
Based on the scaling laws in \eqref{eqn:ridgescale}, we seek a solution in the ridge layer of the form
\begin{eqnarray}
f&=&\wo^{1/2}f_0+\wo^{1/4}f_1+f_2+\ldots, \label{expandf}\\ 
\chi&=&\wo^{1/2}\chi_0+\wo^{1/4}\chi_1+\chi_2+\ldots, \label{expandchi}\\ 
F&=&\wo F_0 + \wo^{3/4}F_1+\wo^{1/2}F_2+\ldots. \label{expandforce}
\end{eqnarray}
In posing these series, we are motivated by numerical solutions to seek expansions in powers of  $\Delta^{-1/4}$; this can be justified  \emph{a posteriori}. Our procedure is to substitute these expansions into \eqref{fvk2:nd} and \eqref{fvk1int:nd} and solve at sequential orders, considering  $x > 0$ and $x < 0$ separately at each stage due to the discontinuity at $x=0$. 

\subsection{Leading order}
The leading problem is at $O(\Delta)$ for which \eqref{fvk2:nd} and \eqref{fvk1int:nd} become
\begin{eqnarray}
&&\frac{\upd^2 f_0}{\upd x^2}-k^4\chi_0\left[f_0 -\mathrm{sgn}(x)\right] = -\frac{k^4 F_0}{2 \pi}\mathrm{sgn}(x)f_0, \label{ridgeleading1} \\
&&\frac{\upd^2\chi_0}{\upd x^2}+\frac{1}{2}f_0\left[f_0 -2\,\mathrm{sgn}(x)\right] = 0,  \label{ridgeleading2}
\end{eqnarray}
valid for both $x < 0$ and $x > 0$. The jump conditions \eqref{eqn:chidiscty} at $x = 0$ read
\beq
[f_0]_{-}^{+}=2, \quad [f_0']_{-}^{+}=0, \quad [\chi_0]_{-}^{+}=\frac{F_0}{\pi}, \quad [\chi_0']_{-}^{+}=0. \label{ridgeleading5}
\eeq

To show that $F_0 = 0$, we follow the approach adopted in the indentation of a shell by a plane~\cite{audoly10}. After multiplying \eqref{ridgeleading1} by $f_0'$ and \eqref{ridgeleading2} by $k^4 \chi_0'$ respectively, we subtract and integrate to yield the first integral
\beq
0 = \mathrm{sgn}(x)\frac{k^4 F_0}{4 \pi}f_0^2 + \frac{1}{2}\left[\left(\frac{\upd f_0}{\upd x}\right)^2-k^4\left(\frac{\upd \chi_0}{\upd x}\right)^2\right]-k^4\frac{f_0}{2}\left[f_0-2\,\mathrm{sgn}(x)\right]\chi_0. \label{leadingridgecons}
\eeq
Evaluating at $x = 0+$ and $x = 0-$ and subtracting, this reduces to $F_0=0$ upon re-arranging using \eqref{ridgeleading5}. Hence the force is not, despite the predicted scaling in \eqref{eqn:ridgescale}, linear in $\wo$. 

\begin{figure}
\centering
\includegraphics[width=12cm]{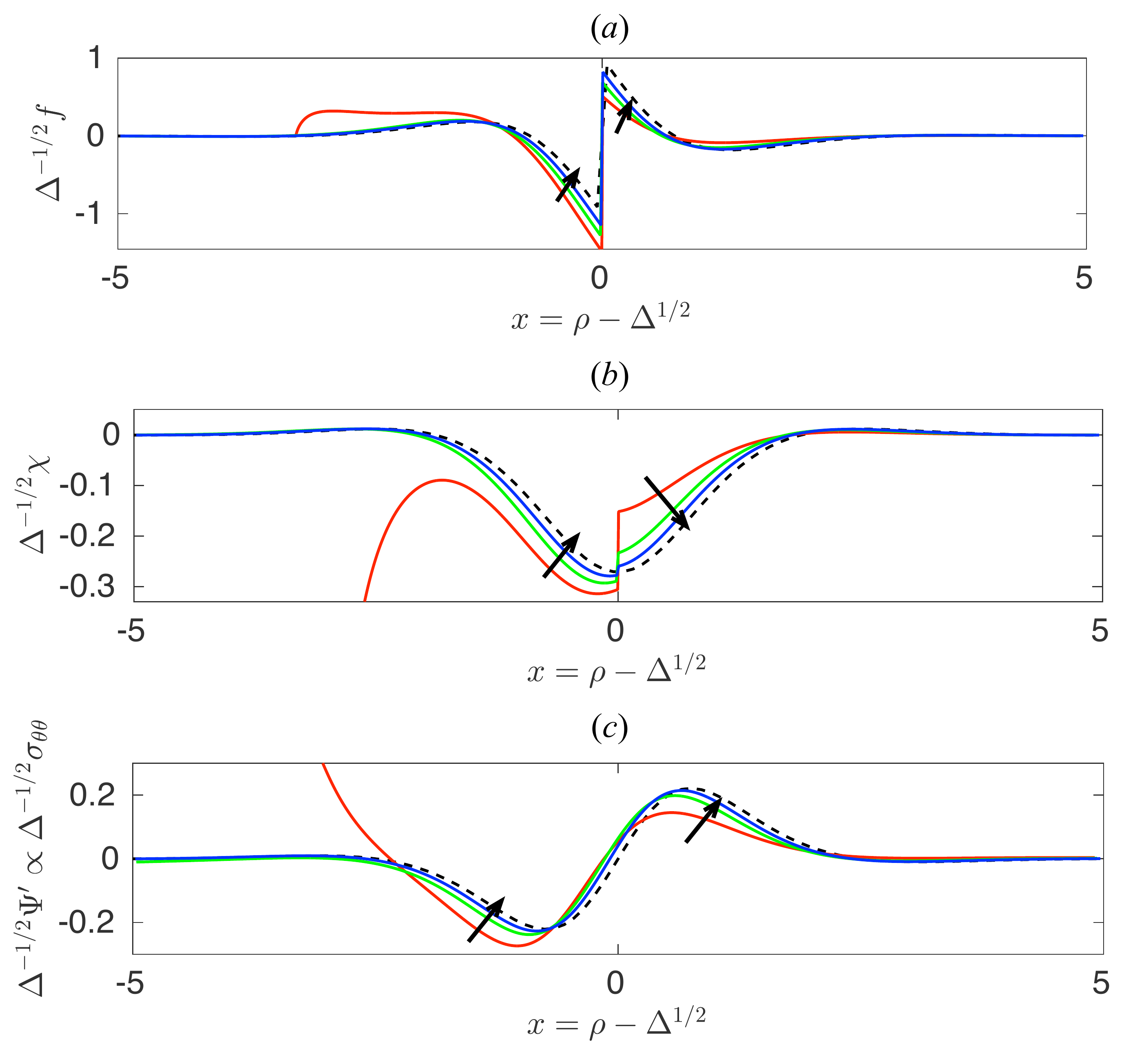}
\caption{Solution in the ridge: comparison between the numerical solution of the full problem (solid curves) and the leading-order asymptotic solution (dashed curves), in terms of the local variable $x = \rho -\Delta^{1/2}$. Numerical results are shown for $\wo=10$, $\wo=100$, and $\wo=1000$, with $\nu = 0.3$. (a), (b): Comparison between $(f_0,\chi_0)$ and $\Delta^{-1/2}(f,\chi)$; the functions $f$ and $\chi$ are related to $W'$ and $\Psi$ by \eqref{defnf} and \eqref{defnchi} (the numerical value of the force $F$ is used to calculate $\chi$, so that the error with the asymptotic result introduces a discontinuity at $x = 0$). (c) The behaviour of the hoop stress $\sqq$, which is related to $\Delta^{-1/2}\Psi'(\rho)$ via $\sqq \propto \Psi'$. In each case, arrows indicate the direction of increasing $\Delta$.}
\label{fig:blvsnums}
\end{figure}

Having shown that $F_0=0$, we deduce that the system \eqref{ridgeleading1}--\eqref{ridgeleading2} has the symmetry
\beq
f_0 \to -f_0,\quad\chi_0\to\chi_0\quad \text{when}\quad x\to-x. \label{ridgeleading6}
\eeq
As such, it is sufficient to solve \eqref{ridgeleading1}--\eqref{ridgeleading2} for $x>0$ with boundary conditions
\beq
f_0(0+)=1,\quad\chi_0'(0+)=0,
\label{ridgeleading7}
\eeq and requiring decay as $x\to+\infty$; this ensures the jump and matching conditions are satisfied. The behaviour for $x<0$ follows by symmetry \eqref{ridgeleading6}.

In fact, with $F_0 = 0$ \eqref{ridgeleading1}  can be shown to be the Euler-Lagrange equation associated with minimizing the total (leading-order) elastic energy of the ridge
\beq
J[f_0,\chi_0] = k^{-1} \int_0^{\infty} \left(\frac{\upd f_0}{\upd x}\right)^2+k^4\left(\frac{\upd \chi_0}{\upd x}\right)^2 \upd x, \label{ridgeenergy}
\eeq 
subject to \eqref{ridgeleading2} and boundary conditions \eqref{ridgeleading7}; the terms in the integral represent the dominant contributions of meridional bending and hoop strain. Minimization of \eqref{ridgeenergy} is precisely the problem considered by Pogorelov~\cite{pogorelov}, who found an approximate analytical solution based on assumptions about the size of terms in \eqref{ridgeleading2}. Here we directly solve the nonlinear ODEs \eqref{ridgeleading1}--\eqref{ridgeleading2} numerically; the solution for $(f_0,\chi_0)$ (obtained using the \textsc{matlab} routine \texttt{bvp4c}) is shown in figure \ref{fig:blvsnums}a,b. Figure \ref{fig:blvsnums} also shows $\Delta^{-1/2}f$ and $\Delta^{-1/2}\chi$ plotted for different values of $\wo$ from the numerical solution of the full problem \eqref{fvk1:nd}--\eqref{nd:bcsinf}. The agreement is very good and, as expected, improves as $\wo$ increases.

\subsection{First order}
\label{firstorder}
The first correction to the leading-order problem above occurs at $O(\Delta^{3/4})$. Equations \eqref{fvk2:nd} and \eqref{fvk1int:nd} become 
%\begin{eqnarray*}
%&&\frac{\upd^2 f_1}{\upd x^2}-k^4(\chi_0f_1+\chi_1f_0-\chi_1) = -\frac{k^4 F_1}{2 \pi}f_0,\\
%&&\frac{\upd^2\chi_1}{\upd x^2}+f_1(f_0-1) = 0,  
%\end{eqnarray*}
%while for $x < 0$ we have
%\begin{eqnarray*}
%&&\frac{\upd^2 f_1}{\upd x^2}-k^4(\chi_0f_1+\chi_1f_0+\chi_1) = \frac{k^4 F_1}{2 \pi}f_0, \\
%&&\frac{\upd^2\chi_1}{\upd x^2}+f_1(f_0+1) = 0. 
%\end{eqnarray*}
\beq
\mathbf{L}(f_1,\chi_1) = \binom{-\frac{k^4 F_1}{2 \pi}\mathrm{sgn}(x)f_0}{0}, \label{ridgefirst1}
\eeq
for both $x > 0$ and $x < 0$, where we have introduced the linear operator
\beq
\mathbf{L}(a,b)\equiv \binom{\frac{\upd^2a}{\upd x^2}-k^4\left[\chi_0a+f_0 b-\mathrm{sgn}(x)b\right]}{\frac{\upd^2 b}{\upd x^2}+a\left[f_0-\mathrm{sgn}(x)\right]}. 
\label{eqn:operator}
\eeq
The boundary conditions are
\beq
[f_1]_{-}^{+}=0, \quad [f_1']_{-}^{+}=0, \quad [\chi_1]_{-}^{+}=\frac{F_1}{\pi}, \quad [\chi_1']_{-}^{+}=0, \label{ridgefirst2}
\eeq
together with decaying conditions at infinity. Note that due to the inhomogeneous right-hand side in \eqref{ridgefirst1}, the Fredholm Alternative Theorem \cite{keener2000} implies that solutions will only exist for a certain value of $F_1$. Furthermore, with this value of $F_1$, we still have the freedom to add to the functions $f_1$ and $\chi_1$ any solution of the corresponding homogeneous problem.
 
To make progress, we note that the operator $\mathbf{L}(\cdot,\cdot)$ can be obtained by differentiating the left-hand side of the leading-order problem \eqref{ridgeleading1}--\eqref{ridgeleading2}. We therefore have that
\beqn
\mathbf{L}(f_0',\chi_0') = \mathbf{0}.
\eeqn
Furthermore, we use \eqref{ridgeleading1}--\eqref{ridgeleading2} with \eqref{ridgeleading7} to show that $f_0''(0\pm)=0$ and $\chi_0''(0\pm) = 1/2$. Recalling that $[f_0']_{-}^{+}=[\chi_0']_{-}^{+}=0$ from \eqref{ridgeleading5}, we have that this solution of the homogeneous problem, $(f_0',\chi_0')$, also satisfies the homogeneous jump conditions
\beqn
[f_0']_{-}^{+}=[f_0'']_{-}^{+}=[\chi_0']_{-}^{+}=[\chi_0'']_{-}^{+}=0.
\eeqn

From the solution of the homogeneous problem, we formulate a solvability condition on $F_1$ by seeking functions $(\phi_1(x),\phi_2(x))$ that satisfy the homogeneous adjoint problem, which reads
\beq
\mathbf{L}(\phi_1,-k^{-4}\phi_2) = \mathbf{0}, \quad [\phi_1]_{-}^{+} = [\phi_1']_{-}^{+}=[\phi_2]_{-}^{+}= [\phi_2']_{-}^{+}=0, \label{ridgefirst3}
\eeq
together with decaying conditions at infinity. The operator $\mathbf{L}(\cdot,\cdot)$ is therefore self-adjoint up to a factor of $-k^{-4}$ in its second argument. The above discussion then allows us to immediately write down a solution of the homogeneous adjoint problem:
\beqn
(\phi_1,\phi_2) = (f_0',-k^4\chi_0').
\eeqn
In particular, the parity of the solutions of the leading-order problem are inherited: $\phi_1$ is even and $\phi_2$ is odd. It is of course possible that other linearly-independent solutions, for example with $\phi_1$ odd and $\phi_2$ even, may exist. Our attempts to find such solutions numerically yielded only trivial solutions and so we assume henceforth  $(f_0',-k^4\chi_0')$ spans the solution space.

Having determined the solution of the homogeneous adjoint problem, we return to the original inhomogeneous problem: dotting \eqref{ridgefirst1} with $(\phi_1,\phi_2)^{\mathrm{T}}$ and integrating over $(-\infty,0-)$ and $(0+,\infty)$, we find (after two integrations by parts) that
\begin{eqnarray*}
\int_{-\infty}^{0-} \frac{k^4 F_1}{2 \pi}f_0\phi_1 \upd x + \int_{0+}^{\infty} -\frac{k^4 F_1}{2 \pi}f_0 \phi_1 \upd x & = & -\left[\phi_1 \frac{\upd f_1}{\upd x}-\frac{\upd \phi_1}{\upd x}f_1 +\phi_2 \frac{\upd \chi_1}{\upd x}-\frac{\upd \phi_2}{\upd x}\chi_1\right] _{0-}^{0+} \\
&& +\:\left(\int_{-\infty}^{0-}+\int_{0+}^{\infty}\right)\binom{f_1}{\chi_1}\cdot\mathbf{L}(\phi_1,-k^{-4}\phi_2) ~\upd x.
\end{eqnarray*}
The expression in square brackets can be simplified using the jump conditions in \eqref{ridgefirst2} and \eqref{ridgefirst3} on $(f_0,\chi_0)$ and $(\phi_1,\phi_2)$, while the first integral on the LHS is a multiple of the second by the parity of $f_0$ and $\phi_1$. Since the integrand on the RHS vanishes by construction, we have
\beq
\frac{F_1}{\pi}\left(\phi_2'(0+)+k^4\int_{0+}^{\infty} f_0 \phi_1 \upd x \right) = 0. \label{ridgefirst6}
\eeq
Taking $(\phi_1,\phi_2) = (f_0',-k^4\chi_0')$, the term in brackets simplifies to $-k^4 < 0$ so that $F_1 = 0$. 

We have seen that the first two terms in the power series of $F$,  \eqref{expandforce}, vanish, i.e.
\beq
F \sim \Delta^{1/2} F_2. \label{forcecomponent}
\eeq
The classic scaling-law of~\cite{pogorelov} leads us to expect that the pre-factor $F_2$ must be non-zero; to evaluate this constant, we must consider the second-order problem --- this is deferred to \S\ref{forcelaw}.

\subsection{The stress in the ridge} 
Having ascertained that $F_1=0$ and that we have a non-zero solution for $(f_0,\chi_0)$ in the absence of a force component at leading order, we can gain a qualitative understanding of the secondary (azimuthal) buckling instability.  The solutions \eqref{mirrorsoln} and \eqref{flatsoln} imply that $\Psi'$ goes from $-F/(2 \pi\rho^2)$ to $+F/(2 \pi\rho^2)$ in moving from the mirror-buckled region to the undeformed region. Using the scaling law $F =O(\Delta^{1/2})$, together with $\rho = O(\Delta^{1/2})$ in the ridge, the net change in $\Psi'$ in moving across the ridge is then $O(\Delta^{-1/2})$. However, the preceding analysis shows that in the ridge itself
\beq
\Psi'(\rho) \sim \Delta^{1/2} \chi_0'(x), \label{ridgestress}
\eeq
which is an order of $\Delta$ larger! Since $\chi_0$ (and $\chi_0'$) decays exponentially at infinity, this leading-order component is not `felt' in the stress beyond the immediate vicinity of the ridge. 

The function $\chi_0'(x)$ has a (negative) minimum  for $x < 0$; see figure~\ref{fig:blvsnums}c, which also shows the rescaled values of $\Psi'$ from the full numerical solutions of the problem. Recalling that the hoop stress $\sqq = \psi' \propto \Psi'\sim\wo^{1/2}\chi_0'(x)$, the existence of a negative minimum in $\chi_0'(x)$ implies the existence of strong, compressive hoop stress in the inner neighbourhood of the ridge. We suggest that this surprisingly strong compression underlies the eventual secondary transition to asymmetrical buckling that is observed in spherical shells under point indentation: for large indentations $\wo\gg1$, the strong compression soon makes it favourable for the ridge to `buckle' out of its axisymmetric shape. Such an explanation has previously been suggested in the context of pressurized spherical shells upon deflation, where the dimpled configuration exhibits a similar instability to form lobes around the ridge; see \cite{knoche2014}.

\subsection{Solution for $(f_1,\chi_1)$} 
Since $F_1=0$,  the apparently inhomogeneous problem \eqref{ridgefirst1}--\eqref{ridgefirst2} is, in fact,  homogeneous.  From the earlier discussion, we can immediately write the solution, up to a constant $\mu_1$, as:
\beq
(f_1,\chi_1) = \mu_1 (f_0',\chi_0').  \label{ridgefirst7}
\eeq
Crucially, the constant $\mu_1$ is not determined by the inner problem at the ridge layer (region 6). Also, a comparison of $(f,\chi)$ computed numerically and the leading-order solution $(f_0,\chi_0)$ suggests that $\mu_1 \neq 0$. Hence we currently have two unknown constants: $\mu_1$ and $A_1$ (the constant in the expansion $W\sim A_1 + \rho^2$ in the mirror-buckled region, \eqref{mirrorsoln}). That these two constants are related can be seen by noting that, when written in terms of the ridge coordinate $x$, the mirror-buckled displacement becomes
\beqn
W \sim (A_1 + \Delta) + 2x \Delta^{1/2} +x^2.
\eeqn
Since the force-balance and compatibility equations \eqref{fvk2:nd} and \eqref{fvk1int:nd} only involve derivatives of $W$, the value of the constant term in $W$ at any point must be determined entirely by the contribution of the slope away from the boundaries $\rho = 0,~\infty$. Integrating  $W'$ (via \eqref{defnf}) over the ridge layer, from the mirror-buckled solution to the undeformed state, i.e.~for $x$ from $-\infty$ to $\infty$ we find
\beqn
A_1 + \Delta \sim -\left(\int_{-\infty}^{0-}+\int_{0+}^{\infty} \right) f(x) \upd x,
\eeqn
Recalling $f\sim f_0+\Delta^{-1/4} \mu_1 f_0'$ with $f_0$ odd, $f_0(0+)=1$ and $f_0(\infty)=0$, we then have that
\beq
A_1 \sim -\Delta+2\Delta^{1/4}\mu_1. \label{constantA1}
\eeq

Having used $W(\infty) = 0$ to obtain an expansion for $A_1$, we require additional information from close to the indenter, $\rho=0$, to evaluate $\mu_1$. This involves better understanding the scaling laws and leading-order solutions that the shell passes through as we decrease $\rho$ away from the ridge. 

\section{Asymptotic structure of the deformation}
\label{solnstructure}

In \S\ref{interiorlayer}, we focused on the interior ridge layer (region $6$ in figure~\ref{sizeofterms1}  at $\rho = \Delta^{1/2}$ and with $O(1)$  width) that accommodates the transition between mirror buckling (with $W \sim A_1 + \rho^2$ and $\Psi \sim F/(2 \pi\rho)$ --- region $5$) and the undeformed solution ($W =o(\Delta^n)
~\forall n,~\Psi \sim -F/(2 \pi\rho)$ --- region $7$). We have already seen that the undeformed solution is uniformly valid for all $\rho \gg \Delta^{1/2}$, while the mirror-buckled solution must break down as  $\rho$ decreases to ensure a finite stress as the indentation point is approached (the final condition in \eqref{nd:bcs0}). It seems reasonable to expect that this breakdown should occur relatively close to the indenter and be accommodated by the introduction of a new interior region.   However, the change in Gauss curvature which we neglected in the mirror-buckled region, measured by the left-hand side of \eqref{fvk2:nd}, scales like $\Psi/\rho \sim F/\rho^2$ (ignoring numerical pre-factors of order unity); while the terms $\rho W'$ and $W'^2$ on the right-hand side both scale like $\rho^2$. Using $F = O(\Delta^{1/2})$, this implies that stretching stresses first become important again when $\rho^4$ is comparable to $\Delta^{1/2}$, i.e.~when $\rho = O(\Delta^{1/8})$.

\begin{figure}[!htb]
\centering
\includegraphics[width=12cm]{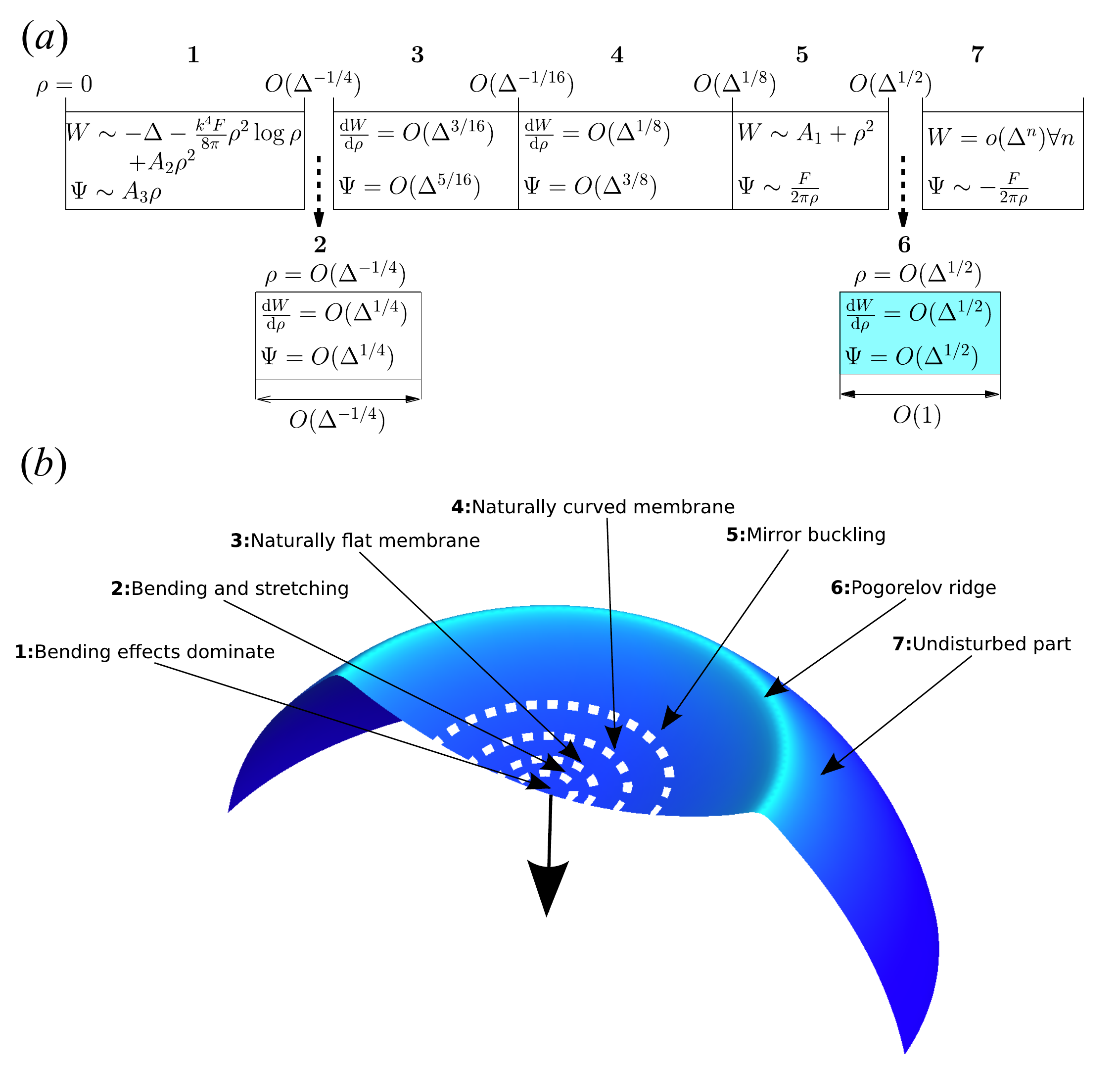}
\caption{The complete asymptotic structure of the solution to \eqref{fvk1:nd}--\eqref{nd:bcsinf} in the limit of large indentations $\Delta \gg 1$ features seven regions in total. (a) The scaling laws for $\Psi$ and $W'$ within each region, together with the size and position of the regions. (b) The physical nature of the balance in each region in a schematic representation of the shell. The classic picture of mirror buckling contains only regions $5,6$ and $7$ but we show that to be consistent with point indentation a further four internal regions are needed.}
\label{asympstructure}
\end{figure}

The breakdown of mirror buckling when $\rho = O(\Delta^{1/8})$ is somewhat surprising and motivates a more careful examination of  the structure of the shell deformation in the limit $\Delta \gg 1$. For clarity, we summarize the results and their physical interpretation in figure~\ref{asympstructure}. However,  for now, we note only that the seven regions (and their approximate positions) correspond reasonably well to the different regions that we earlier identified numerically (see the vertical dashed lines in figure~\ref{sizeofterms1}). We shall proceed by filling in the details of the scaling laws and the balances between terms in each region; we use the numerical solutions as a guide as well as a check of our predictions. 

\subsection{Solution in region $4$}

When $\rho = O(\Delta^{1/8})$, mirror buckling breaks down; at this point, we have $W' \propto \rho = O(\Delta^{1/8})$ and $\Psi \propto \wo^{1/2}/\rho= O(\Delta^{3/8})$ (since $F = O(\Delta^{1/2})$). Balancing the left-hand side of the compatibility equation \eqref{fvk2:nd} with the remaining terms, of size $O(\Delta^{1/4})$, we find that the width of the region in which these scalings hold is $O(\Delta^{1/8})$: unlike the ridge layer (region $6$), region $4$ is not an interior layer but has a width comparable to its radial position.

Guided by these scalings, we introduce the local coordinate $\tilde{\rho}=\rho/ \Delta^{1/8}$ of order unity. The expansions in the neighbouring mirror-buckled region then become:
\beqn
\frac{\upd W}{\upd \rho} \sim 2 \Delta^{1/8}\tilde{\rho}, \quad \Psi \sim \Delta^{3/8}\frac{F_2}{2 \pi \trho}.
\eeqn
It is then natural to introduce (in region 4) the functions $\tf$ and $\tchi$:
\beq
\frac{\upd W}{\upd \rho} = \Delta^{1/8}\left[2\tilde{\rho}+\tf(\trho)\right], \quad \Psi = \Delta^{3/8}\left[\frac{F_2}{2 \pi \trho}+\tchi(\trho)\right]. \label{middlelayer1}
\eeq 
Writing $\tf \sim \tf_0 + o(1)$, $\tchi \sim \tchi_0 + o(1)$ in \eqref{fvk2:nd} and \eqref{fvk1int:nd} we obtain at leading-order
\begin{eqnarray}
 &&\tf_0\left(\tchi_0+\frac{F_2}{2\pi\trho}\right)+\trho\tchi_0=0, \label{middlelayer2} \\
 && \trho\frac{\upd }{\upd \trho}\left[\frac{1}{\trho}\frac{\upd }{\upd \trho}\left(\trho\tchi_0\right)\right]=-\trho\tf_0-\frac{1}{2}\tf_0^2. \label{middlelayer3}
\end{eqnarray}
Note that the bending term does not enter in \eqref{middlelayer2}. The shell remains in a membrane state and its natural curvature enters the force-balance via the $\trho\tchi_0$ term (this is $\rho \Psi$ in the original dimensionless variables). In the compatibility equation, rotations are moderate with a balance between all terms. This is verified by the numerical results in figure~\ref{sizeofterms1} for the particular case $\Delta = 10^3$. However, we know that this region cannot extend all the way to the origin, since the solution as $\rho \to 0$ exhibits a different balance with bending becoming dominant.

Matching from region $4$ into the mirror-buckled region requires
\beq
\tf_0 \to 0, \quad \tchi_0 \to 0 \quad \mathrm{as} \quad\trho \to \infty. \label{middlelayer4}
\eeq
Later, we will demonstrate that the solution around $\rho = 0$ (region $1$) breaks down at $\rho = O(\Delta^{-1/4})$, where $W' = O(\Delta^{1/4})$ and $\Psi = O(\Delta^{1/4})$. However, we cannot yet be sure whether region $4$ borders this internal region, or whether further regions are needed. If a further region is needed, we expect this to bring the solution `close' to the behaviour at the edge of the indenter region, i.e.~$W' = O(\Delta^{1/4})$ and $\Psi = O(\Delta^{1/4})$. We shall therefore make the mild assumption that, in either case, in the region bordering region $4$ we should have $W' \gg \Delta^{1/8}$ and $\Psi \ll \Delta^{3/8}$. The matching conditions into the bordering region then become
\beq
\tf_0 \to \infty, \quad \tchi_0 \sim -\frac{F_2}{2 \pi \trho}\quad \mathrm{as} \quad \trho \to 0. \label{middlelayer5}
\eeq With these boundary conditions, the leading-order problem \eqref{middlelayer2}--\eqref{middlelayer3} can be solved analytically. To see this, note that $\tf_0$ can be eliminated from \eqref{middlelayer3} using \eqref{middlelayer2} \cite{chopin08,vella15} to give
\beqn
\trho\frac{\upd }{\upd \trho}\left[\frac{1}{\trho}\frac{\upd }{\upd \trho}\left(\trho\tchi_0\right)\right]=\frac{\trho^2}{2}\left[1-\left(\frac{F_2/(2\pi\trho)}{F_2/(2\pi\trho)+\tchi_0}\right)^2\right].
\eeqn
Upon making the change of variables \cite{bhatia68,chopin08,vella15}
\beq
\eta = \left(\frac{4 F_2}{\pi}\right)^{-1/2}\trho^2, \quad \tphi_0(\eta) = 1+\frac{\tchi_0(\trho)}{F_2/(2\pi\trho)}, \label{middlelayer6}
\eeq
the matching conditions \eqref{middlelayer4} and \eqref{middlelayer5} become $\tphi_0(0) = 0$ and $\tphi_0(\infty) = 1$, while \eqref{middlelayer6} becomes
\beqn
\frac{\upd^2 \tphi_0}{\upd \eta^2}=1-\tphi_0^{-2}.
\eeqn This may be integrated (with the assumption that $\tphi_0 \nearrow 1$ and $\tphi_0'\searrow 0$ as $\eta \to \infty$) to give
\beq
\frac{\upd \tphi_0}{\upd \eta}=2^{1/2} \frac{1-\tphi_0}{\tphi_0^{1/2}}. \label{middlelayer7}
\eeq
A further integration, using $\tphi_0(0) = 0$, gives the implicit solution
\beq
\atanh\left(\tphi_0^{1/2}\right)-\tphi_0^{1/2} = 2^{-1/2}\eta. 
\label{middlelayer75}
\eeq
%For each $\tphi_0 \in (0,1)$, this equation can be used to find the value of $\eta > 0$. 
Using \eqref{middlelayer2} and \eqref{middlelayer6},  $\tf_0$ and $\tchi_0$ are given in terms of the parameter $\tphi_0 \in (0,1)$ by
\beq
\trho = \left(\frac{4 F_2}{\pi}\right)^{1/4}\eta^{1/2}, \quad \tf_0 = \trho \frac{1-\tphi_0}{\tphi_0},\quad\tchi_0 = \frac{F_2}{2\pi\trho}\left(\tphi_0 -1\right).\label{middlelayer8}
\eeq
Plots of $\tf_0(\trho)$ and $\tchi_0(\trho)$ may be easily generated from \eqref{middlelayer8}. However, what is most valuable about the solution in \eqref{middlelayer75} is that we may determine the limits of its validity.

\subsection{Break-down of region $4$}
As $\eta \searrow 0$, we also have $\tphi_0 \searrow 0$ and so \eqref{middlelayer75} shows that 
\beqn
\tphi_0 \sim \left(\frac{9}{2}\right)^{1/3}\eta^{2/3} \quad \mathrm{as} \quad \eta \to 0.
\eeqn
Recasting this in terms of $\tf$ and $\trho$, as in \eqref{middlelayer8}, we then have that
\beqn
\tf_0 \sim \left(\frac{8 F_2}{9 \pi}\right)^{1/3}\trho^{-1/3}, \quad \quad \tchi_0 \sim -\frac{F_2}{2 \pi \trho}+\left(\frac{3 F_2}{8 \pi}\right)^{2/3}\trho^{1/3}\quad \mathrm{as} \quad \trho \to 0.
\eeqn
In our original variables (see \eqref{middlelayer1}), we see that as $\rho\to0$ within region $4$
\beq
\frac{\upd W}{\upd \rho} \sim \Delta^{1/6}\left(\frac{8 F_2}{9 \pi}\right)^{1/3}\rho^{-1/3}, \quad \Psi \sim \Delta^{1/3}\left(\frac{3 F_2}{8 \pi}\right)^{2/3}\rho^{1/3}. \label{middlelayer9}
\eeq
However, the bending term neglected in the leading-order equation \eqref{middlelayer2} scales like $W'/\rho \sim \Delta^{1/6}\rho^{-4/3}$ (ignoring pre-factors of order unity). Meanwhile the remaining terms $\rho\Psi$ and $\Psi W'$ in the force-balance equation scale like $\Delta^{1/3}\rho^{4/3}$ and $\Delta^{1/2}$, respectively. The bending term balances the $\rho\Psi$ term first, when $\rho = O(\Delta^{-1/16})$, so that region $4$ is $\Delta^{-1/16} \ll \rho \lesssim \Delta^{1/8}$ (as in figure~\ref{asympstructure}a). 

\subsection{Solution in region $3$}

At the inner boundary of region $4$, $\rho = O(\Delta^{-1/16})$ and so, from \eqref{middlelayer9},  $W' = O(\Delta^{3/16})$ and $\Psi= O(\Delta^{5/16})$ as we enter region $3$ (confirming that $W' \gg \Delta^{1/8}$ and $\Psi \ll \Delta^{3/8}$ here, as assumed before \eqref{middlelayer5}). A standard balance argument on the compatibility equation shows that the layer width consistent with these magnitudes is $O(\Delta^{-1/16})$, and so we introduce rescaled (`hat') quantities of order unity
\beqn
\rho = \Delta^{-1/16}\hrho, \quad W' = \Delta^{3/16}\hf, \quad \Psi = \Delta^{5/16}\hchi.
\eeqn
Upon expanding $\hf \sim \hf_0 +o(1)$ and $\hchi \sim \hchi_0 + o(1)$, \eqref{fvk2:nd} and \eqref{fvk1int:nd} reduce, at leading-order, to
\begin{eqnarray}
 && -\hf_0 \hchi_0 = -\frac{F_2}{2 \pi}, \label{region3eqn1} \\
 && \hrho\frac{\upd }{\upd \hrho}\left[\frac{1}{\hrho}\frac{\upd }{\upd \hrho}\left(\hrho\hchi_0\right)\right]=-\frac{1}{2}\hf_0^2. \label{region3eqn2}
\end{eqnarray}
Remarkably, these equations are equivalent to the (geometrically nonlinear) membrane equations for a naturally flat sheet \cite{chopin08}: neither the bending nor natural curvature terms appear and the behaviour is instead dominated by the nonlinear rotation terms ($\Psi W'$ and $W'^2$ in the original variables). This is consistent with the numerical results shown in figure~\ref{sizeofterms1}.

%Written in terms of $\hrho$, the behaviour at the inner edge of region $4$, i.e.~\eqref{middlelayer9}, can be written
%\beqn
%W' \sim \Delta^{3/16}\left(\frac{8 F_2}{9 \pi}\right)^{1/3}\hrho^{-1/3}, \quad \Psi \sim \Delta^{5/16}\left(\frac{3 F_2}{8 \pi}\right)^{2/3}\hrho^{1/3}.
%\eeqn 
The condition that the solution in region $3$ matches into the solution at the inner edge of region $4$, given by \eqref{middlelayer9}, may be written
\beq
\hf_0 \sim \left(\frac{8 F_2}{9 \pi}\right)^{1/3}\hrho^{-1/3}, \quad \hchi_0 \sim \left(\frac{3 F_2}{8 \pi}\right)^{2/3}\hrho^{1/3} \quad \mathrm{as}\quad \hrho \to \infty.
\label{eqn:region3out}
\eeq 
To obtain conditions at the inner edge of region $3$, we again assume that as $\hrho\to0$ the appropriate scalings are `close' to $W' = O(\Delta^{1/4})$ and $\Psi = O(\Delta^{1/4})$, which we know must hold at the outer edge of region $1$. In particular, we anticipate that $W' \gg \Delta^{3/16}$ and $\Psi \ll \Delta^{5/16}$ there, so that 
\beqn
\hf_0 \to \infty, \quad \hchi_0 \to 0 \quad \mathrm{as}\quad \hrho \to 0.
\eeqn 

With these boundary conditions, the leading-order problem \eqref{region3eqn1}--\eqref{region3eqn2} can be solved by eliminating $\hf_0$ from the first equation and using a similar change of variables to \eqref{middlelayer6} used in region $4$, and elsewhere \cite{bhatia68,chopin08,vella15}. We find that the full solution is precisely \eqref{eqn:region3out}, i.e.
\beqn
\hf_0 = \left(\frac{8 F_2}{9 \pi}\right)^{1/3}\hrho^{-1/3}, \quad \hchi_0 = \left(\frac{3 F_2}{8 \pi}\right)^{2/3}\hrho^{1/3}.
\eeqn

\subsection{Break-down of region $3$}
In terms of the original variables, the leading-order solution in region $3$ may be written
\beqn
W' \sim \Delta^{1/6}\left(\frac{8 F_2}{9 \pi}\right)^{1/3}\rho^{-1/3}, \quad \Psi \sim \Delta^{1/3}\left(\frac{3 F_2}{8 \pi}\right)^{2/3}\rho^{1/3}.
\eeqn 
This first breaks down when the bending term neglected in the leading-order force balance (equation \eqref{region3eqn1}) becomes important (the natural curvature terms remain small). As before, this scales like $W'/\rho\sim\Delta^{1/6}\rho^{-4/3}$ (ignoring pre-factors), while the retained term $\Psi W'$ scales like $\Delta^{1/2}$. Equating these gives $\rho = O(\Delta^{-1/4})$, so that the balance in region $3$ can only hold for $\Delta^ {-1/4} \ll \rho \lesssim \Delta^{-1/16}$, as indicated in figure~\ref{asympstructure}.

\subsection{Leading-order problem in region $2$}

Evaluating the solution as we leave region $3$ yields the updated scalings $W' =  O(\Delta^{1/4})$ and $\Psi =  O(\Delta^{1/4})$ in region $2$; again the assumption that $W' \gg \Delta^{3/16}$ and $\Psi \ll \Delta^{5/16}$ is vindicated. The compatibility equation \eqref{fvk2:nd} implies a region width  $O(\Delta^{-1/4})$ for a consistent balance, and so we introduce rescaled variables
\beqn
\rho = \Delta^{-1/4}\rho^{*}, \quad W' = \Delta^{1/4} f^{*}, \quad \Psi = \Delta^{1/4} \chi^{*}.
\eeqn
Expanding $f^{*} \sim f^{*}_0 +o(1)$ and $\chi^{*} \sim \chi^{*}_0 +o(1)$ then, to leading-order, \eqref{fvk2:nd} and \eqref{fvk1int:nd} become
\begin{eqnarray*}
 && k^{-4}\rho^{*}\frac{\upd }{\upd \rho^{*}}\left[\frac{1}{\rho^{*}}\frac{\upd }{\upd \rho^{*}}\left(\rho^{*}f^{*}_0\right)\right]-f^{*}_0\chi^{*}_0 = -\frac{F_2}{2 \pi},  \\
 && \rho^{*}\frac{\upd }{\upd \rho^{*}}\left[\frac{1}{\rho^{*}}\frac{\upd }{\upd \rho^{*}}\left(\rho^{*}\chi^{*}_0\right)\right]=-\frac{1}{2}{f^{*}_0}^2. 
\end{eqnarray*}
The vertical force-balance now involves a balance between bending and in-plane stretching, which is also suggested by the numerical results (figure~\ref{sizeofterms1}). Unfortunately, the presence of the  bending term here prevents us from making further analytical progress. To predict the inner boundary of region $2$, we turn instead to the solution valid in the immediate neighbourhood of the origin, region 1.

\subsection{Solution in region $1$}
As $\rho \to 0$ we anticipate a balance in which bending dominates in-plane stretching --- the boundary conditions \eqref{nd:bcs0} imply that the terms $\rho \Psi$ and $\Psi W'$ must be small (as is confirmed by figure~\ref{sizeofterms1}). Neglecting these terms in \eqref{fvk1int:nd} gives
\beq
k^{-4} \rho\frac{\upd }{\upd \rho}\left[\frac{1}{\rho}\frac{\upd }{\upd \rho}\left(\rho\frac{\upd W}{\upd \rho}\right)\right] = -\frac{F}{2\pi}. \label{region1eqn1}
\eeq Similarly, neglecting the terms $\rho W'$ and $W'^2$ on the RHS of compatibility, \eqref{fvk2:nd}, we have
\beqn
\rho\frac{\upd }{\upd \rho}\left[\frac{1}{\rho}\frac{\upd }{\upd \rho}\left(\rho\Psi\right)\right]=0.
\eeqn
These equations can be solved, subject to the boundary conditions \eqref{nd:bcs0} at $\rho = 0$, to find that
\beq
W \sim -\Delta -\frac{k^4 F}{8 \pi}\rho^2 \log\rho + A_2 \rho^2, \quad \Psi \sim A_3 \rho, \label{smallrhosoln}
\eeq near the indenter, with $A_2$ and $A_3$  constants.

The solution \eqref{smallrhosoln} becomes invalid when the rotation term $\Psi W'$, neglected in \eqref{region1eqn1}, becomes comparable to the bending term $W'/\rho$. This requires that $\Psi \sim 1/\rho$ (ignoring numerical pre-factors), which we know occurs for $\rho \lesssim O(\Delta^{-1/4})$ since this is the scaling law for region $2$.  We therefore also know that $W'\gg\rho$ when the solution breaks down, since $W' \sim \rho \log\rho$ in region $1$. It is possible to show that a balance between bending and stretching with these conditions gives  $W' = O(\Delta^{1/4})$ and $\Psi = O(\Delta^{1/4})$, and hence $\rho \sim 1/\Psi = O(\Delta^{-1/4})$. These are precisely the scaling laws of region $2$; we conclude that region $1$ matches directly into region $2$ (and not into some intermediate region). This completes the final piece of the deformation structure, hinted at by figure~\ref{sizeofterms1} and summarized in figure~\ref{asympstructure}. 

\subsection{Evaluation of $\mu_1$\label{sec:evalMu}}
We are finally in a position to evaluate $\mu_1$ in terms of the force component $F_2$. In region $4$, we  integrate the slope once using the definition of $\tf$ (see \eqref{middlelayer1}). Matching the constant with that in the mirror-buckled region (region $5$), we have that in region $4$
\beqn
W \sim A_1 +\Delta^{1/4}\left(\trho^2 -\int_{\trho}^{\infty} \tf_0(t)\upd t\right).
\eeqn
Evaluating this at $\trho = 0$ gives an expansion for the solution at the outer edge of region $3$ when written in terms of $\trho$. Making use of the expansion \eqref{constantA1} for $A_1$, this can be written as 
\beq
W \sim -\Delta +\Delta^{1/4}\left(2\mu_1 -\int_{0}^{\infty} \tf_0(t)\upd t\right).
\label{eqn:Wend}
\eeq

Alternatively, consider the contribution to the constant term in $W$ due to integrating the slope $W'$ away from $\rho = 0$ where  $W(0)=-\Delta$. In region $1$ we have $W \sim -\Delta -\frac{k^4 F}{8 \pi}\rho^2 \log\rho$ for $\rho$ up to order $O(\Delta^{-1/4})$ (see figure~\ref{asympstructure}); since $F = O(\Delta^{1/2})$, this region contributes an $O(\log\Delta)$ change in $W$. In region $2$, we have $W' = O(\Delta^{1/4})$ for a region of width $O(\Delta^{-1/4})$: this region contributes a further $O(1)$ change to $W$. Similarly, region $3$ can only contribute an $O(\Delta^{1/8})$ change in $W$. When we reach the outer edge of region $3$, we must have that $W = -\Delta + O(\Delta^{1/8})$. Comparing with \eqref{eqn:Wend}, we see that the bracketed term multiplying $\Delta^{1/4}$ in \eqref{eqn:Wend} must vanish, i.e.
\beqn
\mu_1 = \frac{1}{2}\int_{0}^{\infty} \tf_0(t)\upd t.
\eeqn  In other words, since the only contribution at $O(\Delta^{1/4})$ to $W$ comes from integrating the slope over region $4$ and the mirror-buckled region $5$, the boundary conditions at $\rho=0,\infty$ imply these contributions must identically cancel.

Eliminating $\tf_0$ in favour of $\tphi_0$ using \eqref{middlelayer8} we have
\beqn
\int_{0}^{\infty} \tf_0(t)\upd t = \left(\frac{F_2}{\pi}\right)^{1/2}\int_{0}^{\infty} \left[\tphi_0(\eta)^{-1}-1\right]\upd \eta = \left(\frac{F_2}{\pi}\right)^{1/2}\int_{0}^{1} 2^{-1/2}\tphi_0^{-1/2}\upd \tphi_0,
\eeqn
where the last equality follows from changing variables using \eqref{middlelayer7}. Evaluating the last integral, we arrive at the remarkably simple expression
\beq
\mu_1 = \left(\frac{F_2}{2\pi}\right)^{1/2}, \label{mu1value}
\eeq and hence, using \eqref{constantA1}, that the constant $A_1$ in the mirror-buckled solution satisfies
\beqn
A_1 \sim -\Delta+2\left(\frac{F_2}{2\pi}\right)^{1/2}\Delta^{1/4}.
\eeqn 

\section{Force-law}
\label{forcelaw}
The only remaining unknown constant is the pre-factor $F_2$ in the force law $F \sim \Delta^{1/2}F_2$. To determine $F_2$ we return to the ridge layer (region $6$). Recall from \S\ref{interiorlayer} that in this region $f$ and $\chi$ are related to $W'$ and $\Psi$ by \eqref{defnf} and \eqref{defnchi} respectively, while the functions themselves have first-order expansions 
\beq
f \sim \Delta^{1/2}f_0 + \Delta^{1/4}\mu_1 f_0', \quad \chi \sim \Delta^{1/2}\chi_0 + \Delta^{1/4}\mu_1 \chi_0'. \label{newexpandfchi}
\eeq
Here, the functions $(f_0,\chi_0)$ are odd/even respectively and satisfy the leading-order problem \eqref{ridgeleading1}--\eqref{ridgeleading2} with boundary conditions \eqref{ridgeleading7}. As in the case of indentation by a plane~\cite{audoly10}, we expect to determine $F_2$ as a result of a solvability condition on a higher-order problem.

\subsection{Second order}
Following on from the first-order problem in \S\ref{interiorlayer}\ref{firstorder}, the next problem occurs at $O(\Delta^{1/2})$ for which the governing equations \eqref{fvk2:nd} and \eqref{fvk1int:nd} reduce to
\beqn
\mathbf{L}(f_2,\chi_2) = \binom{k^4 \mu_1^2 f_0' \chi_0'-k^4 x f_0\chi_0-\frac{\upd f_0}{\upd x}-\frac{k^4 F_2}{2 \pi}\mathrm{sgn}(x)f_0}{\frac{1}{2}x f_0^2-\frac{\upd \chi_0}{\upd x}-\frac{1}{2}\mu_1^2 f_0'^2}, 
\eeqn
for both $x > 0$ and $x < 0$, with the operator $\mathbf{L}(\cdot,\cdot)$ as defined for the first-order problem, \eqref{eqn:operator} . The boundary conditions ensuring continuity of $W'$ and $\Psi$ (and  first derivatives) at this order are
\beq
[f_2]_{-}^{+}=0, \quad [f_2']_{-}^{+}=2, \quad [\chi_2]_{-}^{+}=\frac{F_2}{\pi}, \quad [\chi_2']_{-}^{+}=0, \label{ridgesecond2}
\eeq
together with decaying conditions at infinity for matching. From the  discussion in \S\ref{interiorlayer}\ref{firstorder}, we see that the homogeneous adjoint problem has  solution $(\phi_1,\phi_2)=(f_0',-k^4\chi_0')$. Since $f_0'$ is even and $\chi_0'$ is odd, to formulate a solvability condition it is natural to first decompose the inhomogeneous right-hand side above into its parts that are odd and even when dotted with $(\phi_1,\phi_2)^{\mathrm{T}}$. We write
\beq
\mathbf{L}(f_2,\chi_2) = \binom{k^4 \mu_1^2 f_0' \chi_0'}{-\frac{1}{2}\mu_1^2 f_0'^2}+\binom{-k^4 x f_0\chi_0-\frac{\upd f_0}{\upd x}-\frac{k^4 F_2}{2 \pi}\mathrm{sgn}(x)f_0}{\frac{1}{2}x f_0^2-\frac{\upd \chi_0}{\upd x}},
\label{ridgesecond1}
\eeq
%where in each term we have grouped components of opposite parity. 
dot \eqref{ridgesecond1} with $(f_0',-k^4\chi_0')^{\mathrm{T}}$, integrate over $(-\infty,0-)$ and $(0+,\infty)$ and add the two results.

The contribution from the first bracketed term in \eqref{ridgesecond1} vanishes (since the integrand is constructed to be odd) along with the dependence on $\mu_1$. In contrast, the contributions from the second bracketed term both reinforce upon adding. Making use of the first-integral \eqref{leadingridgecons} and \eqref{ridgeleading1} (with $F_0 = 0$), this simplifies to
\beqn
\left(\int_{-\infty}^{0-} + \int_{0+}^{\infty}\right)\binom{\phi_1}{\phi_2} \cdot \mathbf{L}(f_2,\chi_2)\upd x = \int_{0+}^{\infty}\left[2k^4\chi_0-\left(f_0'^2 -k^4\chi_0'^2\right)\right] \upd x -2f_0'(0+) +\frac{k^4 F_2}{2 \pi}.
\eeqn
As in the first-order problem, we can integrate by parts twice to shift the operator $\mathbf{L}(\cdot,\cdot)$ onto $(\phi_1,\phi_2)^{\mathrm{T}}$, picking up boundary terms in the process from the jump conditions \eqref{ridgefirst3} and \eqref{ridgesecond2}. Upon re-arranging, we have the solvability condition  
\begin{eqnarray*}
F_2 = \pi k^{-4}\int_{0+}^{\infty} \left(f_0'^2 -k^4\chi_0'^2 -2 k^4 \chi_0\right) \upd x. 
\end{eqnarray*}

Due to our original non-dimensionalization, the expression for $F_2$ is not universal. The solution $(f_0,\chi_0)$ to the leading-order problem  \eqref{ridgeleading1}--\eqref{ridgeleading2} depends on the Poisson ratio $\nu$ (via the constant $k$) in a non-trivial way. This can be scaled out with the change of variables
\beqn
x = k^{-1} \eta, \quad f_0 = 1-\tilde{h}_0, \quad \chi_0 = -k^{-2}\tilde{g}_0,
\eeqn
which transforms the leading-order equations to the shallow shell hinge equations of~\cite{libai}:
\beq
\frac{\upd^2 \tilde{h}_0}{\upd \eta^2} +\tilde{h}_0\tilde{g}_0 = 0, \quad \frac{\upd^2 \tilde{g}_0}{\upd \eta^2} +\frac{1}{2}\left(1-\tilde{h}_0^2\right) = 0,
\eeq
in which $k$ no longer appears. The boundary conditions \eqref{ridgeleading7} and decaying conditions become
\begin{eqnarray*}
&& \tilde{h}_0(0+) = 0,\quad \tilde{g}_0'(0+) = 0, \\
&& \tilde{h}_0 \to 1, \quad \tilde{g}_0 \to 0, \quad \mathrm{as} \quad \eta \to \infty.
\end{eqnarray*}
The solvability condition \eqref{ridgesecond3} can then be re-written as
\beq
F_2 = 2\pi k^{-3}\int_{0+}^{\infty} \frac{1}{2}\left(\tilde{h}_0'^2 -\tilde{g}_0'^2\right) +\tilde{g}_0 \upd \eta. \label{ridgesecond3}
\eeq
The integral can be evaluated numerically as $1.6674\ldots$. Up to numerical error, our result is thus equivalent to the force-law of~\cite{libai}, which in our notation reads
\beqn
F \approx  (1.67) 2\pi k^{-3}\Delta^{1/2}. \label{libaiforce}
\eeqn
This is achieved in \cite{libai}  using the principle of stationary potential energy to relate the work done by the indenter to the elastic energy of the shell, which is dominated by the contribution from the ridge. The method rests on the crucial assumption that $W(0)$ may be evaluated to leading-order using the mirror-buckled solution, namely $W \sim A_1 + \rho^2$. This is equivalent to our earlier assumption that $A_1$ is given by $-\Delta$ to leading-order, which could only be justified formally by a detailed analysis of the size of the corrections arising from regions $1$--$4$ of figure~\ref{asympstructure}.

\begin{figure}
\centering
\includegraphics[width=14cm]{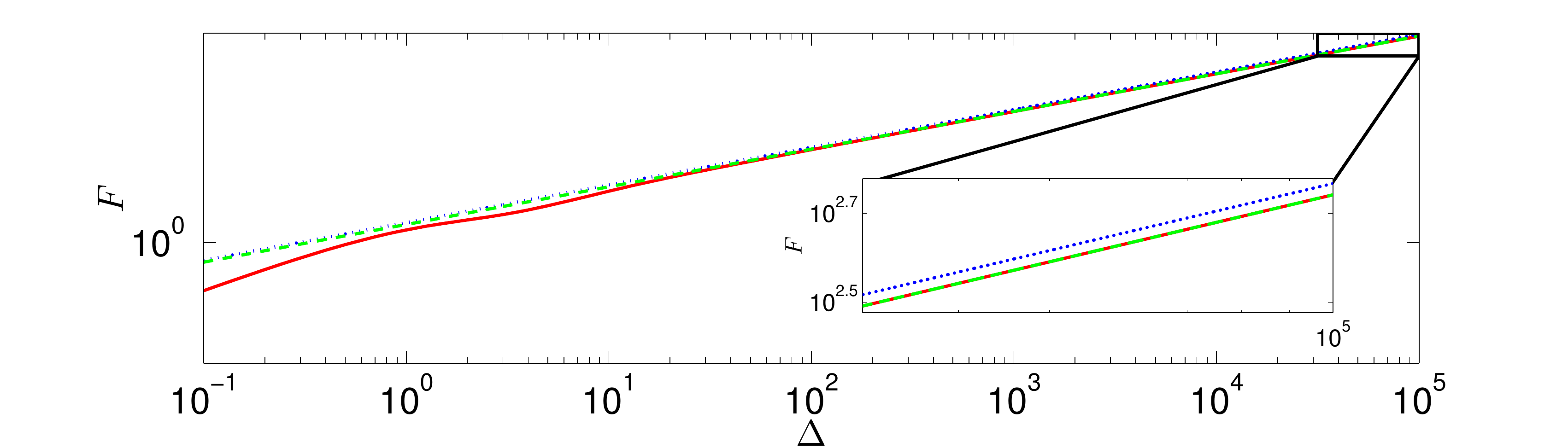}
\caption{The  force--indentation law  obtained by solving the full dimensionless problem \eqref{fvk1:nd}--\eqref{nd:bcsinf} with $\nu=0.3$ (solid curve) compares well with \eqref{forcelawg} (dashed curve) for $\wo\gg1$. Also shown is Pogorelov's result, \eqref{forcelawpogo} (dotted curve).}
\label{forcelaw1}
\end{figure}

In summary, we have shown that in the limit of large indentations $\Delta \gg 1$ we have the force-law 
\beq
F \sim F_2 \Delta^{1/2}, \label{forcelawg}
\eeq
where the pre-factor $F_2$ satisfies the solvability condition \eqref{ridgesecond3}. This result improves the accuracy of the Pogorelov law
\beq
F\sim\cpogo\wo^{1/2}.
\label{forcelawpogo}
\eeq 
Here $\cpogo=3\pi J_0/[12^{3/4}(1-\nu^2)]$ where $J_0\approx1.15092$ is a numerical constant, which is determined by an approximate solution of the leading-order problem in the ridge~\cite{pogorelov}. For the particular case of $\nu = 0.3$, we can compute $F_2 \approx 1.7440$ whilst $\cpogo \approx 1.8488$. A comparison with the numerical result over a range of values of $\Delta$ is given in figure~\ref{forcelaw1} and shows that $F_2$ gives a significantly better approximation of the numerical results  than $\cpogo$ (see inset of figure \ref{forcelaw1}).

\begin{figure}
\centering
\includegraphics[width=12cm]{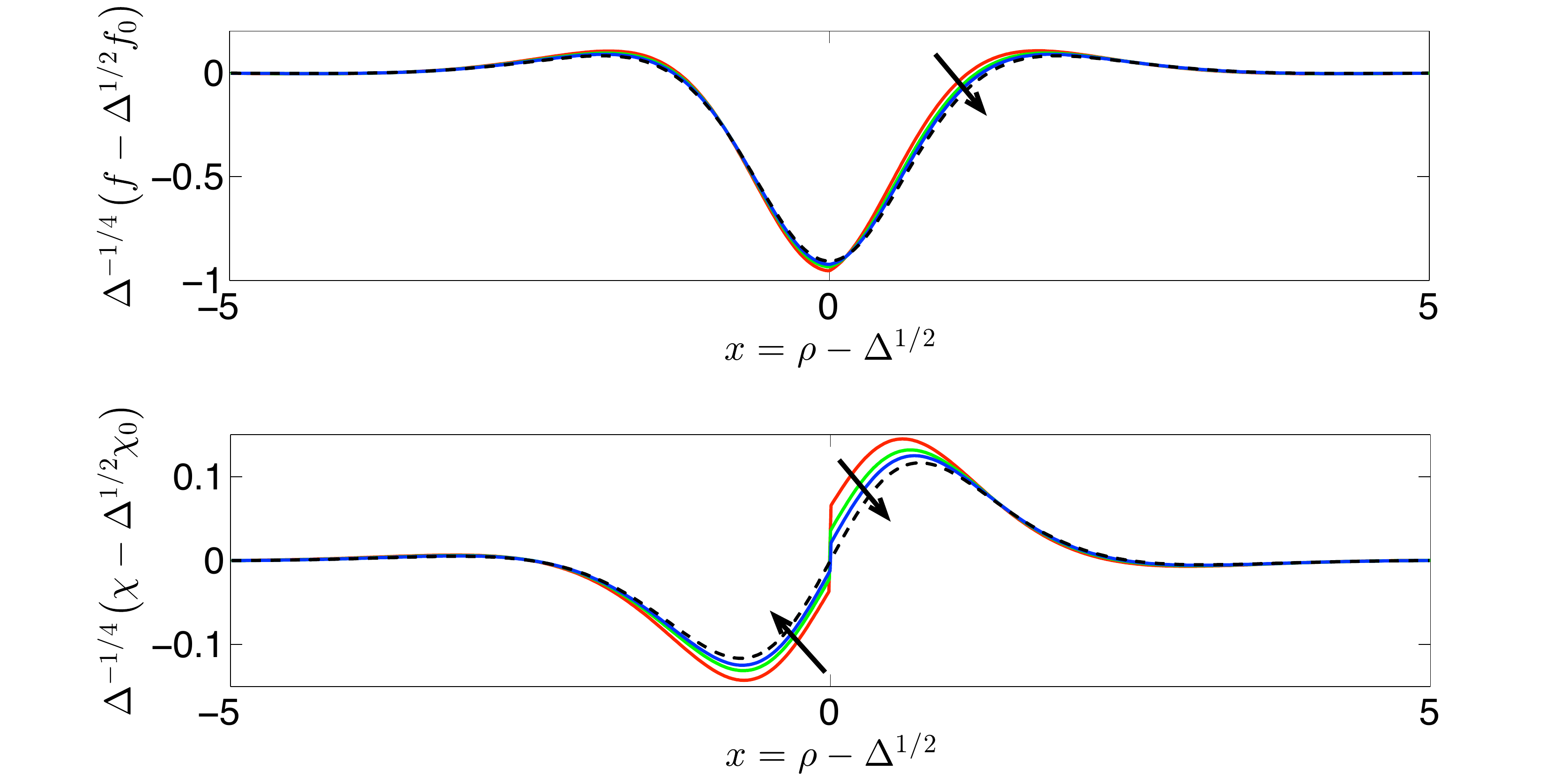}
\caption{Behaviour in the ridge: the errors $\Delta^{-1/4}\left(f-\Delta^{1/2}f_0\right)$ and $\Delta^{-1/4}\left(\chi-\Delta^{1/2}\chi_0\right)$ between the numerical solution $(f,\chi)$ of the full problem and the leading-order asymptotic solution $(f_0,\chi_0)$. Results are shown for $\wo=10^3$, $\wo=10^4$ and $\wo=10^5$ with $\nu = 0.3$ (the numerically determined value of the indentation force, $F$, is used to calculate $\chi$, introducing a small discontinuity at $x = 0$). Also plotted is the approximation $(f_1,\chi_1)=\mu_1 (f_0',\chi_0')$ (dashed curves) with $\mu_1\approx0.5269$ calculated from \eqref{mu1value}. In each plot, arrows indicate the direction of increasing $\Delta$.}
\label{f0chi0numerical}
\end{figure}

The determined value of $F_2$ can also be used to evaluate the constant $\mu_1$, which appears in the first-order problem, by using \eqref{mu1value}. A better check is to compare the numerically-computed error terms $\Delta^{-1/4}\left(f-\Delta^{1/2}f_0\right) $ and $\Delta^{-1/4}\left(\chi-\Delta^{1/2}\chi_0\right)$, which, according to the expansions in \eqref{newexpandfchi}, should be approximated by $\mu_1 (f_0',\chi_0')$. This comparison is shown for a range of large values of $\Delta$ in the case $\nu = 0.3$, when $\mu_1 \approx 0.5269$; see figure~\ref{f0chi0numerical}. The agreement is again excellent.

\subsection{Higher orders}
With a solvability condition at second-order, we can also solve the inhomogeneous problem \eqref{ridgesecond1} with jump conditions \eqref{ridgesecond2}. Crucially, the solution is only determined up to the addition of some multiple $\mu_2$ of $(f_0',\chi_0')$, since this is the solution to the homogeneous problem $\mathbf{L}(f_0',\chi_0')=\mathbf{0}$ with zero jump conditions. Similarly, at third-order a constant $\mu_3$ enters and so on at all higher orders. Like $\mu_1$, these constants are not determined by the inner problem at the ridge, though they appear only at higher orders. Instead, their calculation requires knowledge of asymptotic solutions in regions where they are not currently available, such as the leading-order solution in region $2$. In a sense, our ability to calculate $\mu_1$ here is a stroke of good fortune.

By exploiting linearity and the parity structure of the right-hand side in \eqref{ridgesecond1}, it is possible to decompose the second-order solution in a similar way. The dependence of $\mu_2$ can then be isolated. It turns out that this allows a solvability condition to be obtained for $F_3$ at third-order, where $\mu_2$ cancels in much the same way as $\mu_1$ did above. We find that $F_3 = 0$ so that the next order term in the force-law is at most $O(1)$, i.e.
\beqn
F = \Delta^{1/2}F_2 + O(1).
\eeqn 

The fact that $F_3=0$ explains why the leading-order force law, \eqref{forcelawg}, gives a very good account of the numerical results even at  moderately large values $\Delta \gtrsim 10$ (see figure~\ref{forcelaw1}). As far as we are aware, it is not possible to obtain this bound on the deviation from the leading-order force law by using an energy formulation of the problem. Moreover an analysis of the error between the numerical results and \eqref{forcelawg} suggests that $F_4=0$ also, so that $F = \Delta^{1/2}F_2 + O(\Delta^{-1/4})$. However, we are not able to proceed any further analytically: determining $F_4$ explicitly requires knowledge of at least one of $\mu_2$ or $\mu_3$.

\section{Conclusions}
\label{sectionconclusion}
We have demonstrated that the canonical problem of a spherical shell under point-loading exhibits a broad range of behaviour that has not previously been appreciated. Detailed asymptotic analysis together with numerical results indicate that a seven-region solution structure emerges in the limit of large indentations (figure~\ref{asympstructure}). The key feature of this structure is that it is in contrast with the three-region picture that is usually imagined \cite{pogorelov} and which is observed when contacting with a plane \cite{audoly10}. Crucially, we showed that the stress field associated with mirror buckling is singular as the indenter is approached --- a result that may not be too surprising. What is surprising, however, is that the correction due to bending effects is not localized close to the indenter. Instead, a series of three regions with different scalings are required to match onto mirror buckling. Furthermore, these intermediate regions involve some surprising balances where, for example, the natural curvature and bending stiffness do not play a role.

\begin{figure}
\centering
\includegraphics[width=12cm]{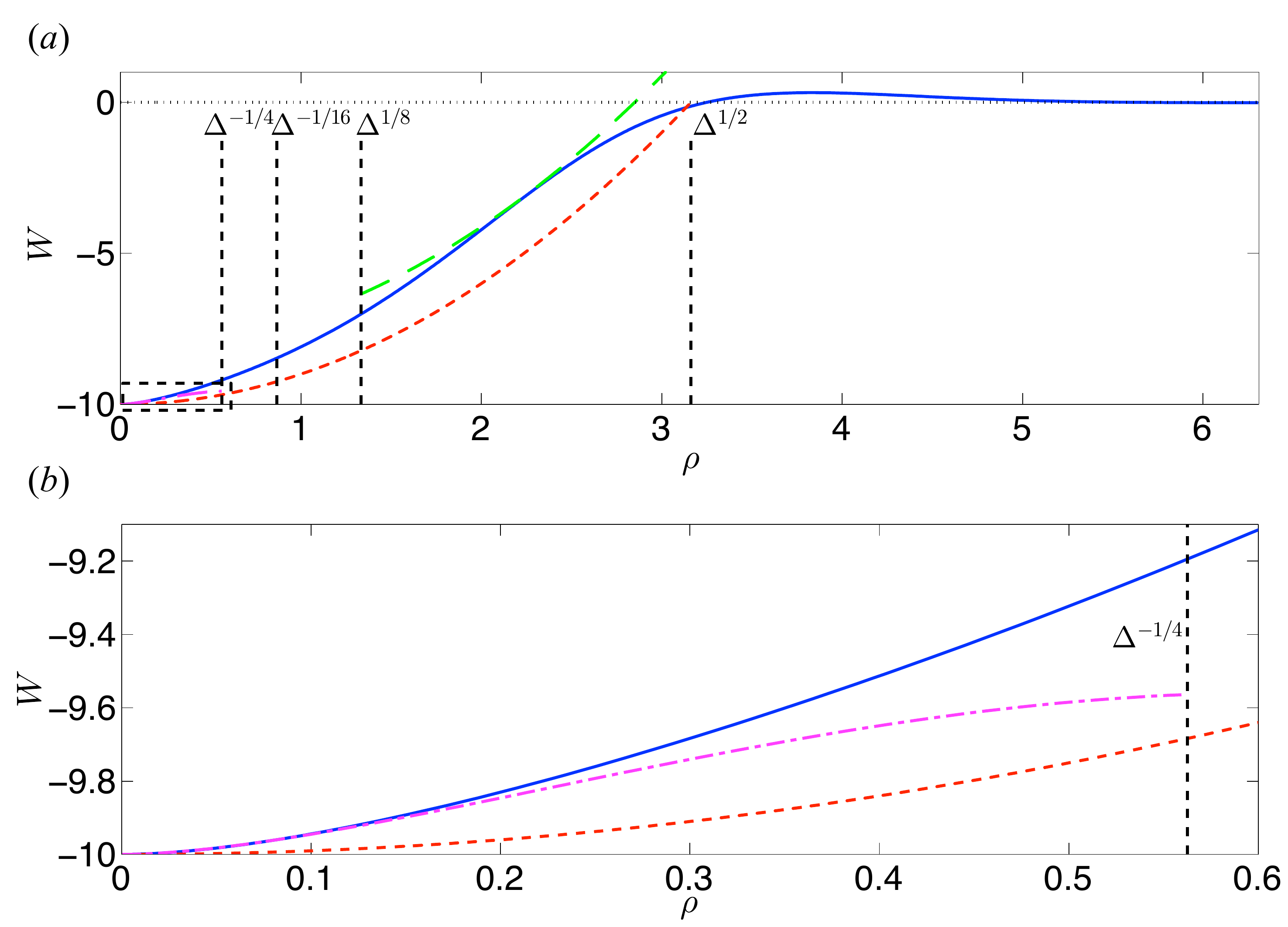}
\caption{({\it a}) The numerically-determined deformation profile $W$ (solid  curves) for $\Delta = 10$ (with $\nu = 0.3$) and ({\it b}) a close-up of the region in the dashed box of ({\it a}) near the indenter. In both plots, the leading-order part of the mirror-buckled solution, $W \sim -\Delta + \rho^2$, is shown (dashed curves). Also displayed are the first correction \eqref{mirrorcorrection} (long dashed curve) and the leading-order behaviour \eqref{smallrhosoln} around $\rho = 0$ (dash-dotted curves). The scaling laws for $\rho$ based on the asymptotic results are shown as vertical dashed lines; the dotted line is at $W = 0$.}
\label{mirrorcorrectionplot}
\end{figure}

While the visual appearance of an indented shell makes mirror buckling seem a plausible model of the deformation, our detailed analysis shows that the internal region where mirror buckling breaks down, of size $\rho=O(\wo^{1/8})$, is in fact an appreciable fraction of the dimple width, $O(\wo^{1/2})$. Indeed, since axisymmetry is lost when $\wo\approx14$ \cite{fitch1968, vaziri08}, this is a significant modification, e.g.~with $\Delta = 10$ (figure \ref{mirrorcorrectionplot}) the non-mirrored portion is around $40\%$ of the inverted region.

A corollary of the asymptotic analysis is the evaluation of the constant $\mu_1$ that appears in the solution of the ridge layer problem. This gives a simple, closed-form expression for the first correction to mirror buckling in terms of the force component $F_2$, namely
\beq
W \sim -\Delta+ \rho^2 + 2\left(\frac{F_2}{2\pi}\right)^{1/2}\Delta^{1/4}. \label{mirrorcorrection}
\eeq
A plot of the numerically-determined displacement $W$ at indentation depth $\Delta = 10$ (before axisymmetry is lost) is shown in figure~\ref{mirrorcorrectionplot}, together with the purely mirror-buckled prediction and its first-order correction, \eqref{mirrorcorrection}. We see that in the region between $\wo^{1/8}$ and $\wo^{1/2}$ where mirror buckling should hold (figure \ref{mirrorcorrectionplot}a), the correction to the mirror-buckled profile is significant. Close to the indenter the bending-dominated prediction, \eqref{smallrhosoln}, offers a clear improvement over mirror-buckling (figure \ref{mirrorcorrectionplot}b). 

Focusing on the Pogorelov ridge layer, we provided an alternative perspective on the well known $F = O(\Delta^{1/2})$ force law, calculating the prefactor $F_2$ using force-balance rather than energy arguments. Balancing bending and stretching effects in the ridge predicted the incorrect scaling law $F = O(\Delta)$, but crucially the stress function $\Psi$ still acts as though this law holds: we showed that the leading-order coefficient $F_0$ in the expansion of the force is zero, though at this order there remains a non-zero solution for $\Psi$ in the ridge. This has the consequence that the magnitude of the hoop stress within the ridge is surprisingly large: we showed that $\sqq = O(\Delta^{1/2})$ in the ridge, unlike the (erroneous) $\sqq = O(1)$ estimate that is actually required to balance the indentation force. This stress rapidly decays outside the ridge in order to match between the mirror-buckled and undeformed regions (see figure \ref{fig:blvsnums}). As indentation progresses and $\Delta$ increases, the compressive azimuthal stress within the ridge therefore grows. Moreover, since the radius of the ridge also increases as the dimple grows but the ridge width remains constant (see Fig.~\ref{asympstructure}), the compressive stress required to buckle it decreases (by analogy with Euler-buckling). From this point of view, secondary buckling is inevitable as indentation progresses.

Finally, we discuss the conditions under which our analysis holds. We have used  equations valid only for shallow shells: the ridge position must be much smaller than the natural curvature of the shell, i.e.~$r_{\mathrm{ridge}} \sim (\delta R)^{1/2}\ll R$. This condition can be satisfied at the same time as the reguirement of large indentation depths $\wo=\delta/h\gg1$ by choosing sufficiently thin shells, $h/R\lll1$. More restrictive is the well-known \cite{fitch1968,vaziri08,vaziri09} result that a spherical shell loses axisymmetry under indentation when $\wo \gtrsim 10$ ($\approx14$ for $\nu=0.3$). This restriction cannot be removed by a careful choice of parameters and, as such, limits the utility of our purely axisymmetric analysis in the limit $\wo\gg1$. Nevertheless, performing this analysis for the axisymmetric equations has yielded new insights into three aspects of the problem: (i) the structure of the solution before instability (see figures~\ref{asympstructure} and~\ref{mirrorcorrectionplot}), (ii) an understanding of why the force law \eqref{forcelawg} works so well for $\wo \gtrsim 10$ when it is formally valid only for $\wo\gg1$, and (iii) the origin of the large compressive hoop stress that ultimately leads to instability.

 \vskip6pt

The research leading to these results has received funding from a Leverhulme Trust Research Fellowship (D.V.), European Research Council under the European Union's Horizon 2020 Programme/ERC grant agreement no. 637334 (D.V.) and the EPSRC (M.G.). D.V.~thanks the Aspen Center for Physics for hospitality during the initial stages of this work and the Zilkha Trust, Lincoln College. We are grateful to Benny Davidovitch and Sebastian Knoche for discussions about this work.

\end{document}